\newif\ifarxiv
\arxivtrue      

\ifarxiv
\documentclass[%
reprint,
superscriptaddress,
nofootinbib,
amsmath,amssymb,
aps,
pre,
notitlepage
]{revtex4-2}
\else
\documentclass[%
reprint,
linenumbers,
superscriptaddress,
nofootinbib,
amsmath,amssymb,
aps,
pre,
notitlepage,
showkeys
]{revtex4-2}
\fi
\usepackage[colorlinks=true, urlcolor=blue, anchorcolor=blue, citecolor=blue,filecolor=blue,linkcolor=blue,menucolor=blue]{hyperref}

\usepackage{subcaption}
\usepackage{graphicx}
\usepackage{dcolumn}
\usepackage{bm}
\usepackage{xcolor}

\begin{document}


\title{The Principle of Isomorphism: A Theory of Population Activity in Grid Cells and Beyond} 

\author{Maoshen Xu}
\altaffiliation{Equal contribution.}
\altaffiliation{Corresponding author.}
\email{maoshenxu@mail.bnu.edu.cn}
\affiliation{School of Systems Science, Beijing Normal University, Beijing, China}
\affiliation{Institute of Natural Sciences, Shanghai Jiao Tong University, Shanghai, China}

\author{Fei Song}
\altaffiliation{Equal contribution.}
\affiliation{School of Psychology, Shanghai Jiao Tong University, Shanghai, China}


\author{Bailu Si}
\altaffiliation{Corresponding author.}
\email{bailusi@bnu.edu.cn}
\affiliation{School of Systems Science, Beijing Normal University, Beijing, China}
\affiliation{Chinese Institute for Brain Research, Beijing, China}

\author{Shanshan Qin}
\altaffiliation{Corresponding author.}
\email{ssqin@sjtu.edu.cn}
\affiliation{Institute of Natural Sciences, Shanghai Jiao Tong University, Shanghai, China}
\affiliation{School of Physics and Astronomy, Shanghai Jiao Tong University, Shanghai, China}

\date{\today}

\begin{abstract}
Neural population activity organizes into low-dimensional manifolds embedded within high-dimensional state spaces, yet the principles governing the topology and geometry of these manifolds remain elusive. Here, we propose the Principle of Isomorphism (PIso), which posits that the topology of a neural manifold is constrained by the mathematical structure of the computational task it supports. We apply this framework to the mammalian grid cell system through two distinct theoretical lenses: an intrinsic neural metric, which requires a locally flat Riemannian structure, and path integration, which requires a compact connected Abelian Lie group structure. We show that these two routes are both sufficient conditions that converge on the same toroidal latent topology, and that they naturally unify within Euclidean space. Using a minimal feed forward network that constrains population activity to a torus with tunable geometry, we find that hexagonal grid fields emerge only in an intermediate geometric regime, becoming diffuse or square-like otherwise. Our work clarifies the separation between three notions—latent topology, extrinsic embedding geometry, and decoded physical geometry—and identifies the topology of the population code as the more invariant consequence of the task structure, while leaving the precise mechanism that selects hexagonal single-cell firing patterns as an open problem.

\end{abstract}

\maketitle


\section{Introduction}



A growing body of work holds that the computational unit of the brain is not the single neuron but the coordinated activity of neural populations \citep{yuste2015neuron,vyas2020computation,kriegeskorte2021neural,perich2025neural}. Large-scale recordings have revealed that while the state space of a neural circuit is high-dimensional, the dynamics are typically confined to low-dimensional manifolds \citep{urai2022large,churchland2012neural,saxena2019towards,langdon2023unifying}. This dimensionality reduction implies the existence of governing laws that constrain neural population activities. However, the theoretical principles that dictate the specific topology and geometry of these manifolds remain an open question.

In this work, we propose a framework to address this question: the Principle of Isomorphism (PIso). PIso posits that the topology of a neural manifold is not arbitrary, but is constrained by the mathematical structure of the computational task it supports. The goal is not to derive every property of population activity, but to make explicit which structural features of the task carry over to which level of the neural code—and to separate this topological consequence from properties of the embedding and of single-cell tuning that need additional assumptions to fix.

We illustrate the framework through a case study of grid cells in the mammalian entorhinal cortex, a system with well-characterized organization at both single-cell and population levels. At the single-cell level, grid cells exhibit hexagonal firing patterns in physical space  \citep{fyhn2004spatial,hafting2005microstructure,sargolini2006conjunctive}. At the population level, the joint activity of cells within the same module has been shown to organize on a toroidal manifold \citep{gardner2022toroidal}. This dual organization makes grid cells an ideal system for asking how the mathematical structure of a computational task is reflected in the structure of its neural representation.

These two features—hexagonal lattice-like firing patterns and toroidal population topology—are part of a broader set of phenomena that any complete theory of grid cells must explain. A recent comprehensive review of normative grid-cell theory \citep{dorrell2026normative} organizes them into four explanatory targets: (P1) hexagonal-lattice tuning curves; (P2) intra-module translational symmetry, where cells in a module share one lattice but are translated copies of one another; (P3) multiple modules with different lattice scales; and (P4) paired conjunctive position--heading-direction cells. From the perspective of PIso, however, these targets are better understood as constraints acting at different levels. Specifically, PIso speaks most directly to the latent topology underlying P2, since a torus is the natural population manifold for a translationally symmetric periodic module. It also provides a natural language for P3, in which distinct modules can be viewed as repeated toroidal latent factors with different scales or embeddings. By contrast, P1 concerns how a latent torus appears in single-cell firing maps after embedding and readout, whereas P4 concerns the circuit-level implementation of motion-induced transformations on the latent manifold.

The functional aspect of grid cells have been centered on \emph{Path integration (PI)} : computing position by integrating self-motion cues and is strongly supported experimentally. PI emphasizes that grid cells are not merely an efficient static code for position, but a spatial representation that must be continuously updated by self-motion \citep{dorrell2026normative}.
A complementary view, which we refer to as the \emph{neural metric (NM)} \citep{moser2008metric,ginosar2023grid,pettersen2024self}, captures the metric requirement for representing position: distances in physical space should be reflected in the intrinsic geometry of the population code. Whether NM is best regarded as a distinct biological objective or as a representational property induced by Euclidean navigation is an open question; here we use it as a formal description of metric structure rather than as an established second biological task.

Historically, theoretical models of grid cells have undergone two major conceptual shifts. The first is a move from mechanistic to normative accounts. Mechanistic models explain how grid fields emerges and how self-motion is integrate to support path integration\citep{burak2009accurate}, whereas normative models explain why grid-like representations should emerge from constraints such as efficient coding, path integration, biological plausibility, or nonlinear readout \citep{stachenfeld2017hippocampus,banino2018vector,cueva2018emergence,sorscher2019unified,whittington2020tolman,gao2021path,xu2023conformal,dorrell2022actionable}. Among normative approaches, conformal-isometry-style (CI) objectives are particularly relevant: they treat grid-cell population activity as a locally conformal embedding of physical space into neural space, accounting for hexagonal response maps through the proportionality between local physical and neural displacements \citep{gao2021path, xu2023conformal, xu2025conformal}. The second shift is from explaining single-cell tuning alone to explaining population-level organization. Population recordings have revealed that neural activities of the same grid module form a coherent toroidal structure \citep{gardner2022toroidal}, and recent numerical experiments suggest that successful path integration may depend more on the coherence of this toroidal manifold than on the precise shape of individual firing maps \citep{schaeffer2022no,schoyen2023coherently}. Together, these shifts raise a central question: how does the hexagonal symmetry of the ``microscopic'' units relate to the ``macroscopic'' topology of the population manifold \citep{sorscher2023unified,schaeffer2023self,pettersen2024self,xu2025conformal}?

\begin{figure*}[t]
  \centering
  \includegraphics[width=0.95\linewidth]{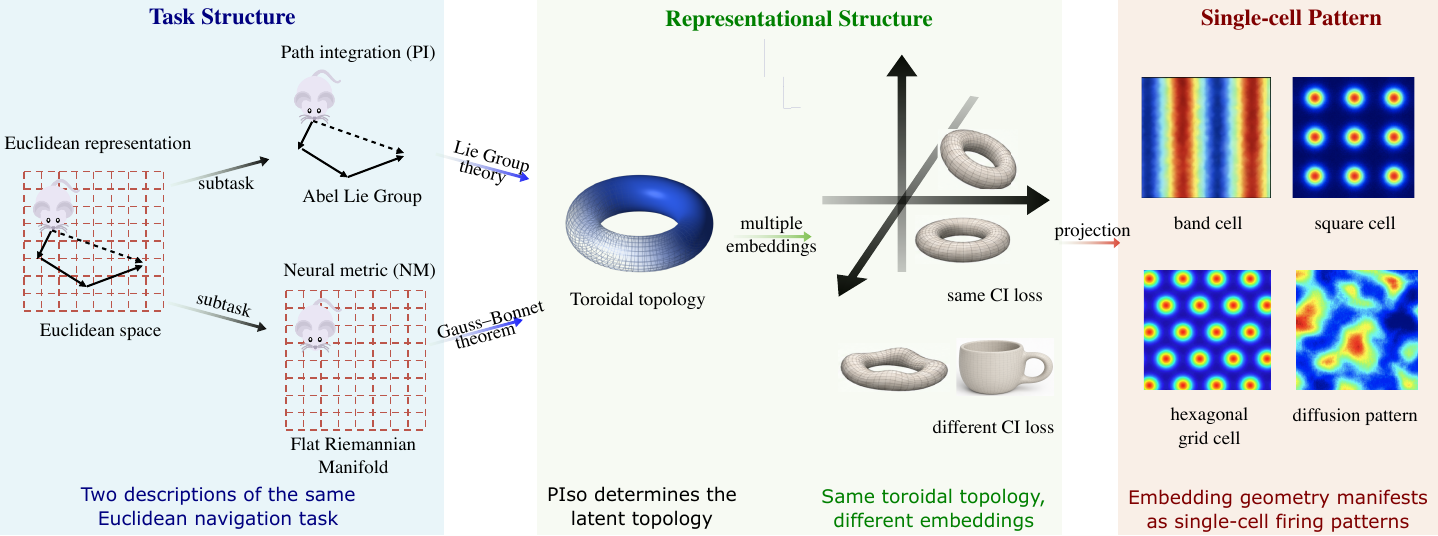}
  \caption{The principle of isomorphism applied to grid cells. Both path integration and neural metric task can be unified as the Euclidean representation task, while each independently determines the toroidal organization of population activity. Moreover, conformal isometry is not a sufficient condition for projecting toroidal population activity into hexagonal firing patterns.}
  \label{fig:frame}
\end{figure*}

We show that PI and NM, viewed through PIso, correspond respectively to a compact connected Abelian Lie group structure and a compact orientable flat Riemannian structure. We treat them as \emph{alternative sufficient routes} to the same latent topology rather than as two jointly necessary biological assumptions. Both routes converge on a toroidal latent code, and the two descriptions are unified at the level of two-dimensional Euclidean space (Fig. \ref{fig:frame}).

While PIso predicts a toroidal topology for the grid-cell population manifold, it does not determine the resulting single-cell firing fields. These firing patterns depend instead on how the torus is embedded in neural activity space (Fig. \ref{fig:frame}). To probe how this latent topology relates to firing fields at the single-cell level, we construct a minimal feedforward model—a topo-constrained network—that explicitly enforces a toroidal latent manifold and exposes its geometry, in particular the torus size, as a control parameter. We find that hexagonal firing patterns emerge robustly only at intermediate torus size, becoming diffuse when the torus is small and square-like when it is large. We use this to argue that toroidal topology together with local CI-style objective is \emph{not} sufficient to guarantee hexagonal single-cell fields across the full range of torus size; an additional geometric or optimization condition is required.

The contributions of this work are threefold (Fig. \ref{fig:frame}):
\begin{itemize}
    \item \textbf{A taxonomy of representational constraints.} We distinguish algebraic constraints from PI, intrinsic metric constraints from NM, and embedding- or readout-level constraints from CI-like local matching as different objects that act at different levels of the neural code.
    \item \textbf{A topology-first result.} Under stated assumptions, both the algebraic PI route and the intrinsic-flat-metric NM route are sufficient to imply a toroidal \emph{latent} topology for the population code; the two routes are unified at the level of Euclidean space.
    \item \textbf{A scope-clarifying negative result.} Within our topo-constrained network, toroidal topology together with CI-like local metric matching does not by itself guarantee hexagonal single-cell fields; hexagonality is restricted to an intermediate torus-size regime, and the precise mechanism that selects it is left as an open problem.
\end{itemize}

In short, PIso clarifies which features of the grid-cell code are forced by task structure (the latent topology) and which require further assumptions (the extrinsic embedding and the hexagonal firing pattern). It does \emph{not} claim that the brain separately optimizes PI and NM, nor that latent topology alone fixes single-cell tuning.

\section{The Principle of Isomorphism}



We introduce a theoretical framework built on the notion that neural population geometry is constrained by task structure. The central concept is the Principle of Isomorphism (PIso), which asserts that the essential structural features of a computational task are preserved in the algebraic and topological organization of the neural population activity.

The framework is defined by three key components:

\textbf{Task Structure}:
This denotes the abstract mathematical structure defining a computational objective. The structure is typically captured by algebraic (groups, rings) or geometric (manifolds, metrics) properties.
\begin{itemize}
    \item \textit{Path Integration (PI)}: The task of integrating displacements $\mathbf{\Delta x}$ to update position $\mathbf{x}$ requires commutative vector addition. This is formally characterized by the continuous translational symmetry of an Abelian Lie group $(\mathbb{R}^2, +)$, a fundamental structure for encoding continuous motion.
    \item \textit{Neural Metric (NM)}: The task of providing an intrinsic spatial map that preserves distances and angles. This is formalized by a Riemannian manifold equipped with a metric tensor. Since the physical world is locally flat, the NM task structure is modeled as a two-dimensional flat Riemannian manifold (zero curvature).
\end{itemize}



\textbf{Representational Structure}: 
This refers to the inherent mathematical organization—topological, geometric, or algebraic—of the manifold $M$ defined by the neural population activity in its high-dimensional state space. For grid cells, the observed manifold is a low-dimensional surface with toroidal topology ($T^2$).

\textbf{Principle of Isomorphism (PIso)}: Subject to essential biological constraints (e.g., compactness, boundarylessness), the Task Structure and Representational Structure must be related by an appropriate structure-preserving map (an isomorphism or homomorphism, depending on the specific structure being preserved). This requires the mapping from the physical variable to the neural activity to preserve the defining operations (e.g., addition, distance).

\textbf{Scope of PIso.}\label{sec:metric_taxonomy}
PIso constrains the \emph{latent} representational structure of a population code, that is, the topology of the manifold on which the population activity lives. To avoid confusion later, we distinguish three notions of ``metric'' or ``geometry'' that recur in this paper:
\begin{itemize}
    \item \emph{Intrinsic metric on the latent manifold.} Distances measured along the manifold itself, independently of how it is embedded in neural activity space. The NM route below acts at this level: it requires a flat intrinsic Riemannian metric.
    \item \emph{Extrinsic geometry of the embedding.} Distances induced by the embedding of the latent manifold into the high-dimensional neural activity space. This is what conformal-isometry-style local matching, used by our numerical model, constrains.
    \item \emph{Decoded geometry in physical space.} The geometry recovered from population activity by a downstream decoder; this depends on both the embedding and the decoder.
\end{itemize}

Throughout the manuscript, ``topology'' refers to a property of the latent manifold and is the object that PIso most directly constrains; ``geometry'' may refer to the intrinsic, extrinsic, or decoded notion above and we make the level explicit when ambiguity could arise.

We emphasize what PIso does and does not claim:
\begin{itemize}
    \item PIso \emph{does} distinguish the minimal latent structure required by a task (its topology) from additional properties of embeddings and readouts.
    \item PIso \emph{does not} prove that the brain separately optimizes PI and NM as two independent biological objectives.
    \item PIso \emph{does not} imply that latent topology alone determines single-cell tuning.
\end{itemize}

PIso connects to a range of concepts from neuroscience and artificial intelligence (AI). In neuroscience, normative theories view neural activity as evolutionarily optimized for ecological demands \citep{attneave1954some,barlow1961possible}, early work on representational geometry emphasized preserving similarity relations \citep{shepard1970second,edelman1998representation}, and recent studies highlight encoding transformations as well as variables \citep{dorrell2022actionable,gao2021path,xu2023conformal}. In AI, related notions appear as inductive biases \citep{wolpert1997no,bengio2013representation}, with geometric deep learning enforcing equivariance to task-defined groups \citep{bronstein2021geometric}, and representational alignment showing convergence of representation across systems solving similar tasks \citep{sucholutsky2023getting}.

\section{Application to Grid Cells}
We apply the PIso framework to the grid cell system, demonstrating how two distinct task structures independently constrain the population activity to the toroidal topology.



\subsection{Neural Metric in the PIso Framework}


\textbf{Task Structure of NM}: 
The NM task requires the neural manifold $M$ to represent a flat physical space, imposing local geometric constraints that, in conjunction with biological constraints, strongly dictate the global topology.

\textbf{Local and Global Constraints}: Under the PIso framework, the representational manifold $M$ must satisfy three necessary constraints:
\begin{itemize}
    \item \textit{Flat Metric:} $M$ must admit a Riemannian metric with zero Gaussian curvature ($K=0$) everywhere, preserving the geometry of a flat, two-dimensional Euclidean plane.
    \item \textit{Compactness:} Neural activity (firing rates) is bounded, excluding infinitely extended manifolds ($\mathbb{R}^2$).
    \item \textit{Boundarylessness:} Continuous spatial representation requires a manifold without boundaries, avoiding discontinuities in the neural code \citep{fiete2008grid}.
\end{itemize}



These conditions strongly restrict the global topology of $M$. The Gauss-Bonnet Theorem provides the fundamental link between the local geometry (curvature, $K$) and the global topology (Euler characteristic, $\chi(M)$) for a compact, boundaryless two-dimensional manifold \citep{needham2021visual}:
\begin{equation}
\int_M K \, dA = 2\pi \chi(M).
\end{equation}
Since the NM constraint requires $K=0$ everywhere (a flat metric), the Euler characteristic must be zero.

The classification of all compact 2D surfaces reveals that only two have an Euler characteristic of zero: the torus and the Klein bottle \citep{hatcher2002algebraic}. Because spatial representation must be orientable—preserving consistent notions of left/right and forward/backward—the Klein bottle is excluded, leaving the torus as the unique solution.

The above analysis shows that the NM task structure, together with compactness, orientability and boundarylessness, restricts the latent grid-cell manifold to a toroidal topology. We emphasize the scope of this statement. The relevant object is the \emph{latent population code} for an effectively boundaryless representation, not the topology of any finite physical environment in which the animal is recorded. Three things should be distinguished: (i) the local task geometry of physical space (locally flat); (ii) the global topology of the latent code (here, a torus); and (iii) the finite arena that is actually sampled by trajectories during experiments or simulations. The conclusion ``the latent code is toroidal'' should be read as a statement about the population manifold under PIso, not as a claim about the physical box. Compactness and boundarylessness should likewise be interpreted at the level of the latent code—reflecting the continuity of the population representation across repeated firing fields—rather than as physical properties of the laboratory environment.

\subsection{Path Integration in the PIso Framework}\label{sec:piso-pi}
The PI task is defined by the algebraic requirement that the neural dynamics integrate displacements via an Abelian Lie group structure. This algebraic route to the torus is closely related to standard arguments in the grid-cell literature \citep{gao2021path,xu2023conformal,whittington2020tolman}; our contribution here is not the route itself, but its alignment with the intrinsic-metric route as an alternative sufficient condition for the same latent topology.

\emph{Algebraic constraint.} Under PIso, the latent representational structure must carry a connected, two-dimensional, compact Abelian Lie group action. According to the classification of Lie groups, the only connected and compact Abelian Lie groups are tori $T^n$ \citep{dwyer1998elementary}. From the PI perspective the toroidal latent topology therefore emerges directly from the requirement that the neural population act as a commutative group for continuous displacement integration.

\emph{Two sufficient routes, not two requirements.} It is important to read the PI and NM analyses correctly. They are two \emph{alternative} sufficient routes to the same toroidal latent topology, not two independent biological objectives that the brain must jointly satisfy. The reason for presenting both is conceptual: the algebraic (PI) and the intrinsic-geometric (NM) descriptions of the same Euclidean navigation task converge on the same latent topology, and this convergence is what we exploit below to argue that the torus is the topological invariant of the task.

\emph{Relation to axis alignment.} Normative work has shown that, among the four standard phenomena of the grid system (hexagonal lattices, intra-module translational symmetry, multimodularity, and conjunctive cells; see \citet{dorrell2026normative}), path integration is specifically what selects translationally symmetric, axis-aligned receptive fields within a module rather than rotated ones \citep{dorrell2022actionable,sorscher2023unified,dorrell2026normative}. Our PI route can be read as the topology-level analogue of this statement: the requirement that the population code carry a compact connected Abelian Lie group action picks out the torus as the latent manifold on which a translationally symmetric, axis-aligned module can live, even though the analysis here does not by itself derive the hexagonal pattern of the single-cell firing fields.




\subsection{Unifying NM and PI: Euclidean representation}

The PIso framework makes explicit that the two routes above are not competing demands but two formal descriptions of a single task structure—the representation of two-dimensional Euclidean space (Fig.~\ref{fig:frame}). We present this unification as a bookkeeping clarification of the relationship between the algebraic and the intrinsic-metric routes.

\textbf{The Common Task Structure}: Two-dimensional Euclidean space ($\mathbb{E}^2$) simultaneously realizes both task requirements:
\begin{itemize}
    \item PI: $\mathbb{E}^2$, viewed as a vector space, supports the Abelian vector addition required for path integration.
    \item NM: $\mathbb{E}^2$, viewed as an inner product space, admits the required flat Riemannian metric.
    \end{itemize}

These two structures are inherently compatible: the group actions (translations and rotations) are isometries (they preserve the inner product structure). This imposes two universal symmetries on the neural manifold:
\begin{itemize}
    \item Homogeneity: No privileged locations; geometric properties are invariant across all positions.
    \item Isotropy: No privileged directions; geometric properties are invariant across all orientations.
\end{itemize}

This unification holds at the representational level as well: the torus ($T^n$) is the only compact manifold that is simultaneously compatible with both the Abelian group structure and a flat Riemannian metric in any dimension.


This observation suggests how the PIso framework might generalize beyond 2D grid cells to other continuous spatial representations \citep{rank1984head,ginosar2021locally,grieves2021irregular,qi2026two}. \textit{Head-direction cells} provide the simplest case: heading is a one-dimensional, flat, and Abelian variable, and therefore corresponds, under PIso, to a 1D torus ($S^1$) latent code. The case of \textit{3D grid cells} is more subtle, and we return to it in the Discussion (Section~\ref{sec:discussion-3d-grid}).

The convergence of algebraic (PI) and intrinsic-geometric (NM) constraints on the same torus is what we take as the central topological consequence of PIso: under the stated assumptions, the latent topology is the level of the population code most directly constrained by task structure, while the extrinsic embedding and the single-cell tuning are not pinned down by this analysis alone. This view is also compatible with classical mechanistic accounts: continuous attractor models that rely on translation-invariant connectivity \citep{skaggs1994model,zhang1996representation,burak2009accurate} can be read as biological implementations of the same Euclidean homogeneity that PIso uses on the task side.



\begin{figure*}[t]
  \centering
  \includegraphics[width=0.8\linewidth]{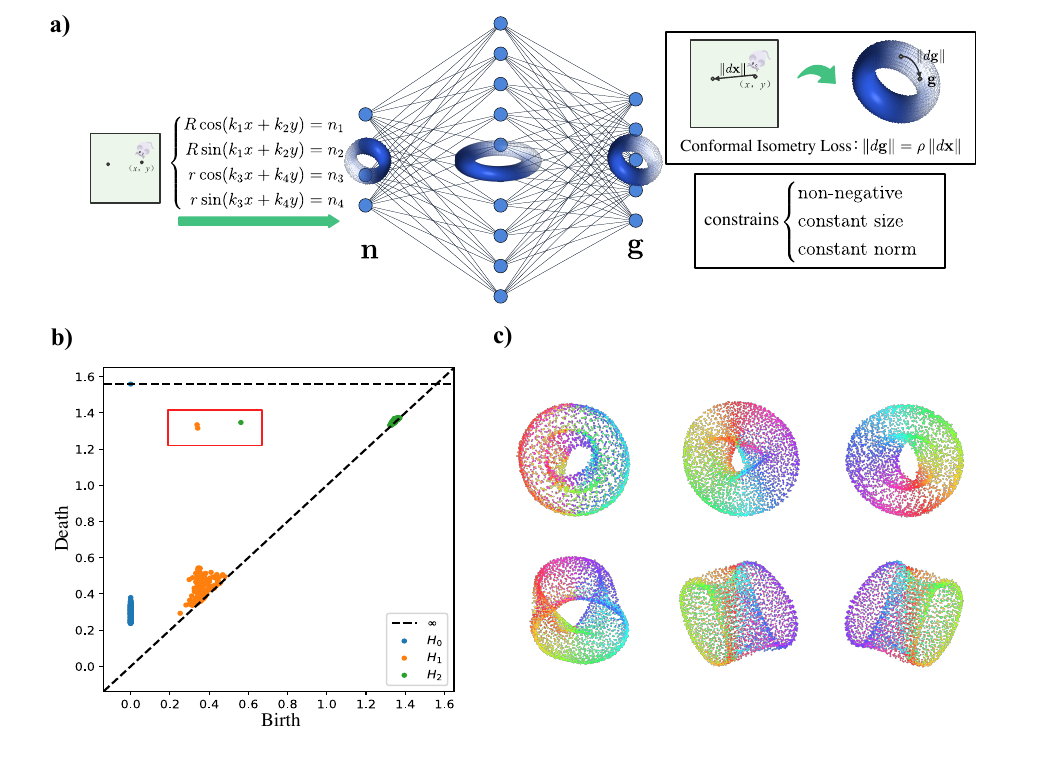}
  \caption{The Topo-Constrained network and toroidal population activity.
    (a) Schematic of the minimal feedforward network used to examine how single-cell firing patterns arise from a prescribed toroidal population topology.
    (b) Persistent homology analysis of the output population activity $\mathbf{g}$ confirms the expected toroidal structure. The persistence diagram shows topological features in dimensions 0, 1, and 2, corresponding respectively to connected components, one-dimensional cycles, and two-dimensional voids.
    (c) Six representative visualizations of the output population activity $\mathbf{g}$. For visualization, high-dimensional population activity was first reduced by PCA and then embedded with UMAP. Points are colored by the value of the first principal component.}
  \label{fig:archi}
\end{figure*}

\begin{figure*}[!t]
  \centering
  \includegraphics[width=\linewidth]{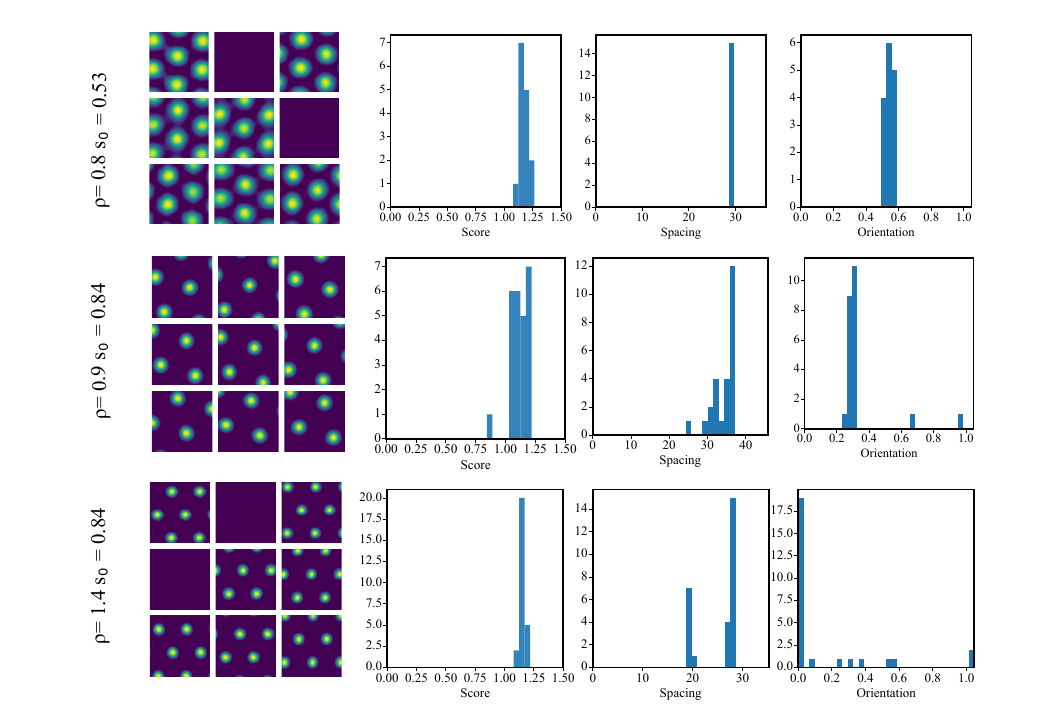}
    \caption{Grid-Like firing patterns under different torus-geometry and mapping-scale parameters. Left: example single-cell firing maps for three representative hyperparameter combinations $(\rho,s_0)$. Right: summary statistics for the corresponding firing maps, including grid score, spacing, and orientation. In the hexagonal-like regime, spacing and orientation are narrowly clustered, consistent with a single grid module; spacing variability becomes more pronounced in the last parameter setting.}
  
  \label{fig:grid_example}
\end{figure*}
\section{Numerical Experiments}


Our theoretical framework PIso identifies toroidal topology as the level of the population code most directly constrained by the navigation task structure, leaving the extrinsic embedding and the single-cell readout under-determined. This raises a natural question: given a toroidal latent code, what additional structure is needed for that code to project onto the hexagonal single-cell lattice that is observed experimentally (Fig. \ref{fig:frame})? We treat the hexagonal pattern as a candidate emergent solution to the geometric problem of mapping a compact, flat toroidal manifold onto a 2D Euclidean plane, but the goal of this section is not to prove that this is the unique mechanism. Instead, we use a minimal feedforward model as a \emph{geometry-controlled probe}: it explicitly fixes the latent topology to a torus and exposes its geometry as a controllable parameter, so that the relationship between latent topology, embedding geometry, and single-cell tuning can be studied in isolation rather than confounded with task learning.

To this end, we constructed the Topo-Constrained Network (TopoCN), which enforces a toroidal latent manifold by construction and lets us vary the geometric parameters of that manifold systematically (Fig.\ref{fig:archi}). We emphasize that TopoCN is not a model of how the brain learns grid cells from a task; it is a probe for testing what aspects of single-cell tuning are determined by torus geometry alone, given the topology.

\subsection{The Topo-Constrained Network (TopoCN)}
As shown in Fig.~\ref{fig:archi}a, the TopoCN is a minimal self-supervised, feedforward network designed to test how toroidal constraints influence single-cell receptive fields.

\textit{Architecture and Constraint Enforcement}. The network maps physical coordinates $\mathbf{x} = (x, y)$ onto a population activity vector $\mathbf{g} \in \mathbb{R}^G$:$$\mathbf{x} \to \mathbf{n}_{\text{torus}} \in \mathbb{R}^4 \to \mathbf{g} \in \mathbb{R}^G$$The intermediate layer $\mathbf{n}_{\text{torus}}$ is explicitly parameterized as a 4D toroidal embedding using learnable rotation and scaling factors [Eq.~\eqref{eq:torus_embed}, Appendix~\ref{sec:num_setup}]. This ensures that the downstream Multi-Layer Perceptron (MLP) operates on a manifold of $T^2$ topology. The output $\mathbf{g}$ is constrained by standard biological requirements: non-negativity ($g_i \ge 0$) and fixed $\ell_2$-norm ($\Vert \mathbf{g} \Vert_2 = 1$), which is supported by experimental observation \citep{gardner2022toroidal, xu2025conformal}.

\textit{Geometric Control Parameters}. Unlike previous models that use capacity-based regularization, our framework controls two critical, task-relevant geometric parameters:
\begin{itemize}
  \item \textbf{Mapping Scale ($\rho$)}: Controls the scale of the mapping between physical displacement and neural displacement, influencing the final grid spacing.
  \item \textbf{Torus Size ($s$)}: Quantifies the compactness of the neural manifold, defined from the mean population-activity vector [Eq.~\eqref{eq:torus_size}, Appendix~\ref{sec:num_setup}].
\end{itemize}
We use a regularization term in the loss function $\mathcal{L}$ to target a specific size $s_0$:
\begin{equation}\label{eq:loss}
    \mathcal{L} = \mathbb{E} \Big[\big(\Vert \Delta \mathbf{g}\Vert -  \rho \Vert \Delta \mathbf{x} \Vert \big)^2
    \exp\!\left(-\tfrac{\Vert \Delta \mathbf{x}\Vert^2}{2\sigma^2}\right) \Big]
    + \lambda (s - s_0)^2,
\end{equation}
where the first term is called Conformal Isometry (CI) loss, and $\Delta \mathbf{x}$ and $\Delta \mathbf{g}$ denote displacement in physical and the corresponding change of representation vector in the neural space, $\sigma$ controls the locality of conformal isometry, $\lambda = 2$ and $s_0 \in [0,1]$ sets the target torus size. In all the numerical results reported in the main text, $G = 32$ was used (See Appendix \ref{sec:num_setup} for more details about the numerical setup).

The CI loss (the first term in $\mathcal{L}$), which encourages the preservation of local angles but allows scaling. This design isolates the CI principle and the toroidal topology, allowing us to test their combined effect on hexagonal firing pattern formation.



\subsection{Results: Emergence and Geometric Dependence}

\subsubsection{A torus-geometry regime in which hexagonal patterns emerge robustly}
We next used the TopoCN introduced above to ask how a prescribed toroidal population topology is expressed at the level of single-cell firing patterns. As shown in Fig.~\ref{fig:archi}b,c, the model constrains the output population activity $\mathbf{g}$ to lie on a toroidal latent manifold, and this topology is verified both by persistent homology and by low-dimensional visualization. We then varied the torus geometry and mapping scale to examine when this fixed latent topology gives rise to grid-like single-cell responses.

Without any explicit constraint favoring periodicity or hexagonal symmetry, TopoCN identifies a regime of torus geometry in which hexagonal-like firing patterns emerge robustly when the toroidal latent manifold is mapped onto single-cell activity. In this regime, the resulting fields show narrow clustering in spacing and orientation, qualitatively consistent with a single grid module (Fig.~\ref{fig:grid_example}; see Appendix~\ref{sec:FF} for additional examples). This indicates that the hexagonal pattern is a natural outcome of mapping a compact, flat $T^2$ manifold onto a 2D plane. Our sweep covers torus size and mapping scale; other geometric degrees of freedom of the flat torus—notably the relative scale and orientation of its two fundamental cycles—together with the readout and optimization, remain open and may support non-hexagonal yet equally functional codes.

At large torus sizes, the firing fields develop heterogeneous spacing within a single network, hinting at a spontaneous tendency toward multiple spatial scales (Fig. \ref{fig:grid_example}). Testing whether this reflects genuine grid modules \citep{schaeffer2023self}—through clustering of spacing, orientation coherence, and phase coverage across seeds—is a natural way to connect torus geometry to the multi-module organization of the entorhinal map.

\begin{figure*}[!t]
  \centering
  \includegraphics[width=0.8\linewidth]{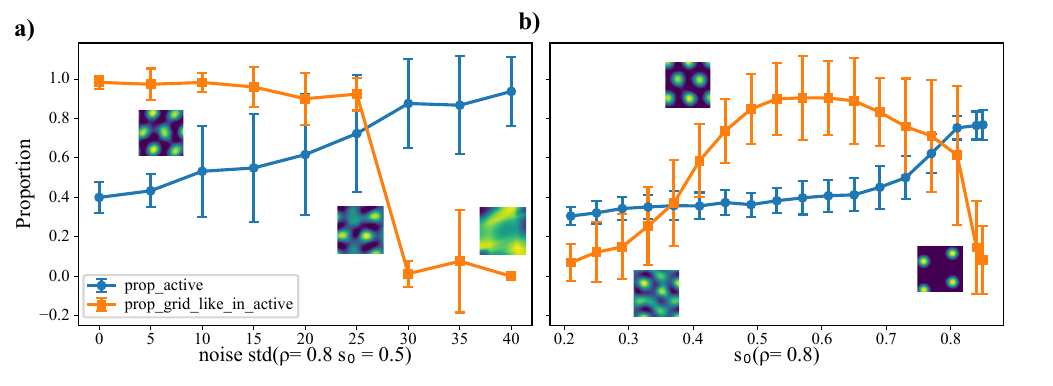}
    \caption{Robustness to noise and dependence on torus size. Proportion of active cells and grid cells as a function of (a) noise level and (b) the torus-size parameter $s_0$. Hexagonal firing patterns are robust to moderate noise but emerge only within an intermediate range of torus sizes.}
  
  \label{fig:grid_statis}
\end{figure*}

\subsubsection{Resilience to noise}

Injecting noise into the population activity, we find that hexagonal firing patterns persist across a wide range of noise levels (Fig.~\ref{fig:grid_statis}a): as noise grows, the network recruits a larger fraction of active neurons to preserve the spatial code. The toroidal-to-hexagonal projection is thus robust to realistic neural variability (see Appendix~\ref{sec:num_setup} for the noise protocol).


\subsubsection {Scope of Conformal Isometry within TopoCN}

Our sweep across the torus-size parameter $s_0$ delineates the regime in which the CI-style local matching loss produces hexagonal fields within TopoCN (Fig.~\ref{fig:grid_statis}b):

\textit{Grid peak}: hexagonal lattices predominantly emerge within an intermediate range of $s_0$.

\textit{Outside this regime}: at large $s_0$ (where the embedding is closest to a maximally flattened torus) the patterns tend toward square-like patterns rather than hexagonal ones, while at small $s_0$ they become diffuse and non-localized.

This dependence on $s_0$ is itself informative: within TopoCN, a toroidal latent topology together with CI-style loss yields hexagonal patterns only at intermediate torus size, not across the full geometric range. 
Hexagonality therefore requires an ingredient beyond topology and local isometry—specifically, a mechanism that selects particular embeddings of the toroidal manifold.
Which factor dominates—network capacity, the relative weighting of the loss terms, mode selection, or optimization bias—is a concrete target for future work.

A possible geometric interpretation is that the torus-size parameter does not directly control grid-pattern formation, but instead biases the embedding of the toroidal manifold in neural activity space. Notably, the CI loss does not uniquely determine this embedding: different embeddings can achieve similar CI values while producing different single-cell firing fields. For example, geometric transformations that preserve local distances on the manifold, such as rotations or translations of the embedding in neural activity space, leave the CI objective essentially unchanged but can alter the firing patterns observed at the single-cell level. From this perspective, the dependence on $s_0$ may arise indirectly through its effect on embedding geometry, suggesting that CI alone is insufficient to uniquely determine grid-cell tuning.

\citet{xu2025conformal} showed that CI can promote a toroidal-to-hexagonal projection by maximally flattening the torus. \citet{dorrell2026normative} further note that the CI-style local matching loss acts as a bandpass filter in Fourier space, in which case the analytic accounts of \citet{sorscher2023unified}—where non-negativity and selection of low-cost wavevectors favor sixfold (hexagonal) combinations—supply the mechanism that picks out hexagonal patterns within the available band. Our results refine this picture: the bandpass/non-negativity mechanism operates within an intermediate range of torus sizes, and connecting it quantitatively to the torus-geometry regime identified here is a natural next step.

\begin{figure*}[!t]
  \centering
  \includegraphics[width=0.8\linewidth]{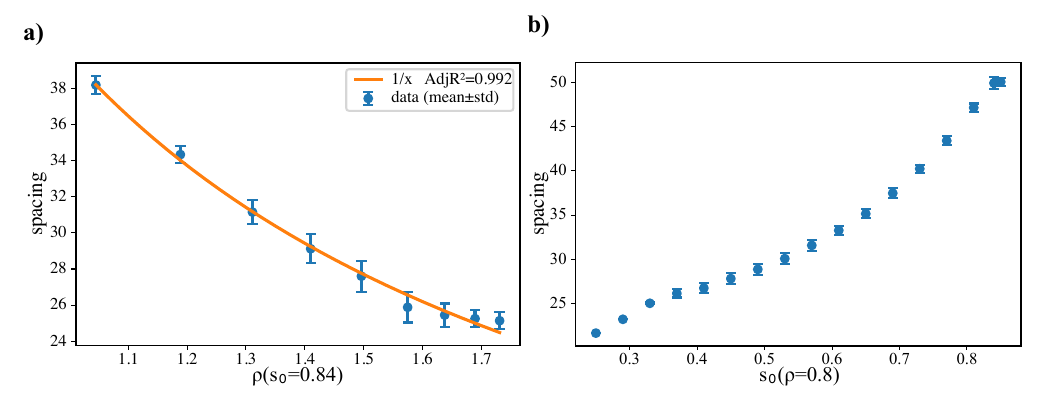}
  \caption{Grid spacing as a function of model parameters. (a) Grid spacing is approximately proportional to $1/\rho$. (b) Grid spacing increases monotonically with the torus-size parameter $s_0$.}
  \label{fig:spacing}
\end{figure*}

\subsubsection{Predictable Control Over Grid Spacing}

The grid spacing—the physical distance between adjacent firing fields—reflects the distance corresponding to one complete traversal around the torus in a given direction. We established a systematic, independent relationship between the two geometric control parameters and the resulting grid cell spacing:

\textbf{Scale Dependence ($\rho$)}: Grid spacing is inversely proportional to the scaling factor $\rho$ (Fig.~\ref{fig:spacing}a) (see Appendix \ref{sec:prac} for details). Since $\rho$ controls the scale of the mapping, this confirms that grid spacing reflects the distance in physical space that corresponds to one complete cycle around the toroidal manifold.

\textbf{Size Dependence ($s_0$)}: Grid spacing increases monotonically with the torus size $s_0$ while holding the scaling factor $\rho$ constant (Fig.~\ref{fig:spacing}b). This novel finding suggests that larger, more distributed toroidal manifolds naturally encode broader spatial representations.

These results show that, within TopoCN, grid spacing is directly controlled by the geometric parameters $\rho$ and $s_0$. We frame the torus size $s_0$ here as an \emph{interpretable control coordinate} for torus geometry, closely related to capacity-style regularizers (Appendix~\ref{sec:cap}) rather than a principled replacement for them; its usefulness is that it organizes the observed firing-field regimes (diffuse, hexagonal, square-like) in a transparent way. The same hexagon-supporting regime is reached spontaneously, without explicit size control, by a recurrent variant of the model that actively performs path integration (Appendix~\ref{sec:PI}), indicating that the effect is not an artifact of the static feedforward network setup.



\section{Relation to prior work}\label{sec:related}

PIso builds on and reorganizes several established lines of grid-cell theory.

\textbf{Geometric perspectives and conformal isometry.}
The conformal isometry (CI) hypothesis \citep{gao2021path,xu2023conformal,xu2025conformal} constrains the \emph{extrinsic} embedding of the population manifold in neural activity space, requiring embedded distances to match physical distances up to a constant scaling. This fixes both geometry and mapping, in tension with the grid-field distortions seen in deformed environments \citep{krupic2015grid,krupic2018local,stensola2015shearing}. PIso instead requires only a flat \emph{intrinsic} metric, which every torus admits; the latent topology is then preserved under stretching or remapping of the embedding, consistent with the persistence of the toroidal topology of the population activity when firing fields distort \citep{gardner2022toroidal} and with the approach to conformal isometry in undistorted regimes \citep{xu2025conformal}.

\textbf{Path integration as a group representation.}
That grid-cell activity preserves displacement algebra \citep{fiete2008grid,sreenivasan2011grid} and commutativity \citep{whittington2020tolman}, and realizes a representation of the planar translation group $(\mathbb{R}^2,+)$ whose compact connected form is a torus \citep{gao2021path,xu2023conformal}, is established. PIso adopts this algebraic route (Sec.~\ref{sec:piso-pi}) and pairs it with the intrinsic-metric route as a second sufficient condition for the same latent topology.

\textbf{Normative and population-level accounts.}
Normative models derive grid structure by optimizing coding or task objectives under biological constraints \citep{sorscher2019unified,banino2018vector,cueva2018emergence,dorrell2022actionable}; analytic versions show that non-negativity selects sixfold lattices \citep{sorscher2023unified} and that path integration aligns module axes \citep{dorrell2022actionable,dorrell2026normative}. A complementary line emphasizes the topology and modularity of the population code over single-cell tuning \citep{schaeffer2022no,schoyen2023coherently,schaeffer2023self,pettersen2024self}. PIso sits at this population level: it identifies the latent torus forced by task structure and is complementary to the single-cell analytic accounts that select the hexagonal lattice within that topology.


\section{Discussion}

We introduced the Principle of Isomorphism (PIso) to relate the mathematical structure of a computational task to the latent topology of the population code that supports it. For grid cells, the algebraic structure of path integration and the flat intrinsic metric of the neural metric are two sufficient routes to the same toroidal latent topology, unified at the level of two-dimensional Euclidean space. A geometry-controlled network (TopoCN) then shows that projecting this torus onto hexagonal single-cell firing patterns requires more than topology alone: robust hexagonal fields appear only in an intermediate regime of torus geometry, with diffuse or square-like fields outside it.

\subsection{Topology is more invariant than embedding geometry}

The central conceptual message of this study is that topology and geometry act at different levels of the code, and that task structure constrains the former more tightly than the latter. Three notions must be kept apart (Sec.~\ref{sec:metric_taxonomy}): the intrinsic metric of the latent manifold, which PIso constrains; the extrinsic geometry of its embedding in neural activity space, which conformal isometry constrains; and the decoded geometry recovered downstream. This separation organizes the experimental observation: the grid patterns distort in deformed environments \citep{krupic2015grid,krupic2018local,stensola2015shearing} while the toroidal population topology persists \citep{gardner2022toroidal}, exactly the dissociation expected if distortion acts on the embedding while topology is the task-determined invariant. It also clarifies decoding---any toroidal embedding in principle retains the information to recover a flat metric through a sufficiently nonlinear decoder, whereas a uniform conformal embedding additionally permits linear decoding \citep{xu2025conformal,pouget2000information,sreenivasan2011grid}. A direct test is to confirm, via persistent homology, that deformed environments leave the toroidal topology intact while distorting firing patterns.

\subsection{3D grid cells and open questions}
\label{sec:discussion-3d-grid}


The preceding analysis treats grid cells as a compact latent representation of Euclidean task structure, where flat metric geometry and Abelian displacement composition are realized together. This raises a natural question for \textit{3D grid cells}. If the task structure to which the animal is adapted is genuinely three-dimensional Euclidean space, then PIso predicts that the optimal latent population code should be a 3-torus ($T^3$). Recent recordings from freely flying bats, however, suggest a different possibility: 3D navigation may be supported by reusing a 2D toroidal grid-cell manifold, aligned to the animal's current plane of motion \citep{qi2026two}. If confirmed, this would indicate that animals do not necessarily implement the theoretical optimum for full volumetric navigation. Rather, because their natural behavior may not require precise path integration and neural metric along all three dimensions, reusing an existing 2D grid architecture may provide an evolutionarily economical and behaviorally sufficient solution. 

Why the latent torus projects so reliably onto hexagonal fields remains largely unknown. Analytic results in which non-negativity and a bandpass filter select sixfold lattices \citep{sorscher2023unified,xu2025conformal,dorrell2026normative} are the natural candidate mechanism, and connecting them to the intermediate torus-geometry regime found here---and to the role of capacity regularizers such as $L_2$ shrinkage---is a clear next step. A topology-first analysis is also silent about the conjunctive position--heading-direction cells that mechanistic continuous-attractor models use to implement path integration.

\subsection{A topology prior for machine learning}

More speculatively, PIso suggests a \emph{topology prior} for representation learning: when a task carries a known symmetry, constraining the latent space to the corresponding manifold---a torus $T^n$ for Euclidean translation, a sphere $S^n$ or rotation group $SO(n)$ for rotational symmetry, a hyperbolic manifold for hierarchies---may provide a useful inductive bias. Whether such priors improve performance or generalization on machine-learning benchmarks is an open empirical question.


\section*{Data Availability Statement}
The code generating the numerical results in this paper is available at GitHub: \url{https://github.com/Neal-Xu/PIso_GridCells}.

\section*{Author Contributions}

\begin{itemize}
\item Maoshen Xu: Conceptualization, methodology, formal analysis, investigation, writing the original draft, and contributed to Software.
\item Fei Song: Led software, contributed to formal analysis, and contributed to writing -- review and editing.

\item Bailu Si: Supervision, project administration, funding acquisition.

\item Shanshan Qin: Supervision, investigation, funding acquisition, and led rewriting and Editing.
\end{itemize}

\section*{Acknowledgments}
We thank Yuxiu Shao for helpful comments on the manuscript. BS was supported by the National Science and Technology Innovation 2030 Major Program of China (2022ZD0205005). SQ was supported by the National Science and Technology Innovation 2030 Major Program of China (2025ZD0218700).

\bibliography{apssamp}

\begin{thebibliography}{53}%
\makeatletter
\providecommand \@ifxundefined [1]{%
 \@ifx{#1\undefined}
}%
\providecommand \@ifnum [1]{%
 \ifnum #1\expandafter \@firstoftwo
 \else \expandafter \@secondoftwo
 \fi
}%
\providecommand \@ifx [1]{%
 \ifx #1\expandafter \@firstoftwo
 \else \expandafter \@secondoftwo
 \fi
}%
\providecommand \natexlab [1]{#1}%
\providecommand \enquote  [1]{``#1''}%
\providecommand \bibnamefont  [1]{#1}%
\providecommand \bibfnamefont [1]{#1}%
\providecommand \citenamefont [1]{#1}%
\providecommand \href@noop [0]{\@secondoftwo}%
\providecommand \href [0]{\begingroup \@sanitize@url \@href}%
\providecommand \@href[1]{\@@startlink{#1}\@@href}%
\providecommand \@@href[1]{\endgroup#1\@@endlink}%
\providecommand \@sanitize@url [0]{\catcode `\\12\catcode `\$12\catcode
  `\&12\catcode `\#12\catcode `\^12\catcode `\_12\catcode `\%12\relax}%
\providecommand \@@startlink[1]{}%
\providecommand \@@endlink[0]{}%
\providecommand \url  [0]{\begingroup\@sanitize@url \@url }%
\providecommand \@url [1]{\endgroup\@href {#1}{\urlprefix }}%
\providecommand \urlprefix  [0]{URL }%
\providecommand \Eprint [0]{\href }%
\providecommand \doibase [0]{https://doi.org/}%
\providecommand \selectlanguage [0]{\@gobble}%
\providecommand \bibinfo  [0]{\@secondoftwo}%
\providecommand \bibfield  [0]{\@secondoftwo}%
\providecommand \translation [1]{[#1]}%
\providecommand \BibitemOpen [0]{}%
\providecommand \bibitemStop [0]{}%
\providecommand \bibitemNoStop [0]{.\EOS\space}%
\providecommand \EOS [0]{\spacefactor3000\relax}%
\providecommand \BibitemShut  [1]{\csname bibitem#1\endcsname}%
\let\auto@bib@innerbib\@empty
\bibitem [{\citenamefont {Yuste}(2015)}]{yuste2015neuron}%
  \BibitemOpen
  \bibfield  {author} {\bibinfo {author} {\bibfnamefont {R.}~\bibnamefont
  {Yuste}},\ }\bibfield  {title} {\bibinfo {title} {From the neuron doctrine to
  neural networks},\ }\href@noop {} {\bibfield  {journal} {\bibinfo  {journal}
  {Nature reviews neuroscience}\ }\textbf {\bibinfo {volume} {16}},\ \bibinfo
  {pages} {487} (\bibinfo {year} {2015})}\BibitemShut {NoStop}%
\bibitem [{\citenamefont {Vyas}\ \emph {et~al.}(2020)\citenamefont {Vyas},
  \citenamefont {Golub}, \citenamefont {Sussillo},\ and\ \citenamefont
  {Shenoy}}]{vyas2020computation}%
  \BibitemOpen
  \bibfield  {author} {\bibinfo {author} {\bibfnamefont {S.}~\bibnamefont
  {Vyas}}, \bibinfo {author} {\bibfnamefont {M.~D.}\ \bibnamefont {Golub}},
  \bibinfo {author} {\bibfnamefont {D.}~\bibnamefont {Sussillo}},\ and\
  \bibinfo {author} {\bibfnamefont {K.~V.}\ \bibnamefont {Shenoy}},\ }\bibfield
   {title} {\bibinfo {title} {Computation through neural population dynamics},\
  }\href@noop {} {\bibfield  {journal} {\bibinfo  {journal} {Annual review of
  neuroscience}\ }\textbf {\bibinfo {volume} {43}},\ \bibinfo {pages} {249}
  (\bibinfo {year} {2020})}\BibitemShut {NoStop}%
\bibitem [{\citenamefont {Kriegeskorte}\ and\ \citenamefont
  {Wei}(2021)}]{kriegeskorte2021neural}%
  \BibitemOpen
  \bibfield  {author} {\bibinfo {author} {\bibfnamefont {N.}~\bibnamefont
  {Kriegeskorte}}\ and\ \bibinfo {author} {\bibfnamefont {X.-X.}\ \bibnamefont
  {Wei}},\ }\bibfield  {title} {\bibinfo {title} {Neural tuning and
  representational geometry},\ }\href@noop {} {\bibfield  {journal} {\bibinfo
  {journal} {Nature Reviews Neuroscience}\ }\textbf {\bibinfo {volume} {22}},\
  \bibinfo {pages} {703} (\bibinfo {year} {2021})}\BibitemShut {NoStop}%
\bibitem [{\citenamefont {Perich}\ \emph {et~al.}(2025)\citenamefont {Perich},
  \citenamefont {Narain},\ and\ \citenamefont {Gallego}}]{perich2025neural}%
  \BibitemOpen
  \bibfield  {author} {\bibinfo {author} {\bibfnamefont {M.~G.}\ \bibnamefont
  {Perich}}, \bibinfo {author} {\bibfnamefont {D.}~\bibnamefont {Narain}},\
  and\ \bibinfo {author} {\bibfnamefont {J.~A.}\ \bibnamefont {Gallego}},\
  }\bibfield  {title} {\bibinfo {title} {A neural manifold view of the brain},\
  }\href@noop {} {\bibfield  {journal} {\bibinfo  {journal} {Nature
  Neuroscience}\ ,\ \bibinfo {pages} {1}} (\bibinfo {year} {2025})}\BibitemShut
  {NoStop}%
\bibitem [{\citenamefont {Urai}\ \emph {et~al.}(2022)\citenamefont {Urai},
  \citenamefont {Doiron}, \citenamefont {Leifer},\ and\ \citenamefont
  {Churchland}}]{urai2022large}%
  \BibitemOpen
  \bibfield  {author} {\bibinfo {author} {\bibfnamefont {A.~E.}\ \bibnamefont
  {Urai}}, \bibinfo {author} {\bibfnamefont {B.}~\bibnamefont {Doiron}},
  \bibinfo {author} {\bibfnamefont {A.~M.}\ \bibnamefont {Leifer}},\ and\
  \bibinfo {author} {\bibfnamefont {A.~K.}\ \bibnamefont {Churchland}},\
  }\bibfield  {title} {\bibinfo {title} {Large-scale neural recordings call for
  new insights to link brain and behavior},\ }\href@noop {} {\bibfield
  {journal} {\bibinfo  {journal} {Nature neuroscience}\ }\textbf {\bibinfo
  {volume} {25}},\ \bibinfo {pages} {11} (\bibinfo {year} {2022})}\BibitemShut
  {NoStop}%
\bibitem [{\citenamefont {Churchland}\ \emph {et~al.}(2012)\citenamefont
  {Churchland}, \citenamefont {Cunningham}, \citenamefont {Kaufman},
  \citenamefont {Foster}, \citenamefont {Nuyujukian}, \citenamefont {Ryu},\
  and\ \citenamefont {Shenoy}}]{churchland2012neural}%
  \BibitemOpen
  \bibfield  {author} {\bibinfo {author} {\bibfnamefont {M.~M.}\ \bibnamefont
  {Churchland}}, \bibinfo {author} {\bibfnamefont {J.~P.}\ \bibnamefont
  {Cunningham}}, \bibinfo {author} {\bibfnamefont {M.~T.}\ \bibnamefont
  {Kaufman}}, \bibinfo {author} {\bibfnamefont {J.~D.}\ \bibnamefont {Foster}},
  \bibinfo {author} {\bibfnamefont {P.}~\bibnamefont {Nuyujukian}}, \bibinfo
  {author} {\bibfnamefont {S.~I.}\ \bibnamefont {Ryu}},\ and\ \bibinfo {author}
  {\bibfnamefont {K.~V.}\ \bibnamefont {Shenoy}},\ }\bibfield  {title}
  {\bibinfo {title} {Neural population dynamics during reaching},\ }\href@noop
  {} {\bibfield  {journal} {\bibinfo  {journal} {Nature}\ }\textbf {\bibinfo
  {volume} {487}},\ \bibinfo {pages} {51} (\bibinfo {year} {2012})}\BibitemShut
  {NoStop}%
\bibitem [{\citenamefont {Saxena}\ and\ \citenamefont
  {Cunningham}(2019)}]{saxena2019towards}%
  \BibitemOpen
  \bibfield  {author} {\bibinfo {author} {\bibfnamefont {S.}~\bibnamefont
  {Saxena}}\ and\ \bibinfo {author} {\bibfnamefont {J.~P.}\ \bibnamefont
  {Cunningham}},\ }\bibfield  {title} {\bibinfo {title} {Towards the neural
  population doctrine},\ }\href@noop {} {\bibfield  {journal} {\bibinfo
  {journal} {Current opinion in neurobiology}\ }\textbf {\bibinfo {volume}
  {55}},\ \bibinfo {pages} {103} (\bibinfo {year} {2019})}\BibitemShut
  {NoStop}%
\bibitem [{\citenamefont {Langdon}\ \emph {et~al.}(2023)\citenamefont
  {Langdon}, \citenamefont {Genkin},\ and\ \citenamefont
  {Engel}}]{langdon2023unifying}%
  \BibitemOpen
  \bibfield  {author} {\bibinfo {author} {\bibfnamefont {C.}~\bibnamefont
  {Langdon}}, \bibinfo {author} {\bibfnamefont {M.}~\bibnamefont {Genkin}},\
  and\ \bibinfo {author} {\bibfnamefont {T.~A.}\ \bibnamefont {Engel}},\
  }\bibfield  {title} {\bibinfo {title} {A unifying perspective on neural
  manifolds and circuits for cognition},\ }\href@noop {} {\bibfield  {journal}
  {\bibinfo  {journal} {Nature Reviews Neuroscience}\ }\textbf {\bibinfo
  {volume} {24}},\ \bibinfo {pages} {363} (\bibinfo {year} {2023})}\BibitemShut
  {NoStop}%
\bibitem [{\citenamefont {Fyhn}\ \emph {et~al.}(2004)\citenamefont {Fyhn},
  \citenamefont {Molden}, \citenamefont {Witter}, \citenamefont {Moser},\ and\
  \citenamefont {Moser}}]{fyhn2004spatial}%
  \BibitemOpen
  \bibfield  {author} {\bibinfo {author} {\bibfnamefont {M.}~\bibnamefont
  {Fyhn}}, \bibinfo {author} {\bibfnamefont {S.}~\bibnamefont {Molden}},
  \bibinfo {author} {\bibfnamefont {M.~P.}\ \bibnamefont {Witter}}, \bibinfo
  {author} {\bibfnamefont {E.~I.}\ \bibnamefont {Moser}},\ and\ \bibinfo
  {author} {\bibfnamefont {M.-B.}\ \bibnamefont {Moser}},\ }\bibfield  {title}
  {\bibinfo {title} {Spatial representation in the entorhinal cortex},\
  }\href@noop {} {\bibfield  {journal} {\bibinfo  {journal} {Science}\ }\textbf
  {\bibinfo {volume} {305}},\ \bibinfo {pages} {1258} (\bibinfo {year}
  {2004})}\BibitemShut {NoStop}%
\bibitem [{\citenamefont {Hafting}\ \emph {et~al.}(2005)\citenamefont
  {Hafting}, \citenamefont {Fyhn}, \citenamefont {Molden}, \citenamefont
  {Moser},\ and\ \citenamefont {Moser}}]{hafting2005microstructure}%
  \BibitemOpen
  \bibfield  {author} {\bibinfo {author} {\bibfnamefont {T.}~\bibnamefont
  {Hafting}}, \bibinfo {author} {\bibfnamefont {M.}~\bibnamefont {Fyhn}},
  \bibinfo {author} {\bibfnamefont {S.}~\bibnamefont {Molden}}, \bibinfo
  {author} {\bibfnamefont {M.-B.}\ \bibnamefont {Moser}},\ and\ \bibinfo
  {author} {\bibfnamefont {E.~I.}\ \bibnamefont {Moser}},\ }\bibfield  {title}
  {\bibinfo {title} {Microstructure of a spatial map in the entorhinal
  cortex},\ }\href@noop {} {\bibfield  {journal} {\bibinfo  {journal} {Nature}\
  }\textbf {\bibinfo {volume} {436}},\ \bibinfo {pages} {801} (\bibinfo {year}
  {2005})}\BibitemShut {NoStop}%
\bibitem [{\citenamefont {Sargolini}\ \emph {et~al.}(2006)\citenamefont
  {Sargolini}, \citenamefont {Fyhn}, \citenamefont {Hafting}, \citenamefont
  {McNaughton}, \citenamefont {Witter}, \citenamefont {Moser},\ and\
  \citenamefont {Moser}}]{sargolini2006conjunctive}%
  \BibitemOpen
  \bibfield  {author} {\bibinfo {author} {\bibfnamefont {F.}~\bibnamefont
  {Sargolini}}, \bibinfo {author} {\bibfnamefont {M.}~\bibnamefont {Fyhn}},
  \bibinfo {author} {\bibfnamefont {T.}~\bibnamefont {Hafting}}, \bibinfo
  {author} {\bibfnamefont {B.~L.}\ \bibnamefont {McNaughton}}, \bibinfo
  {author} {\bibfnamefont {M.~P.}\ \bibnamefont {Witter}}, \bibinfo {author}
  {\bibfnamefont {M.-B.}\ \bibnamefont {Moser}},\ and\ \bibinfo {author}
  {\bibfnamefont {E.~I.}\ \bibnamefont {Moser}},\ }\bibfield  {title} {\bibinfo
  {title} {Conjunctive representation of position, direction, and velocity in
  entorhinal cortex},\ }\href@noop {} {\bibfield  {journal} {\bibinfo
  {journal} {Science}\ }\textbf {\bibinfo {volume} {312}},\ \bibinfo {pages}
  {758} (\bibinfo {year} {2006})}\BibitemShut {NoStop}%
\bibitem [{\citenamefont {Gardner}\ \emph {et~al.}(2022)\citenamefont
  {Gardner}, \citenamefont {Hermansen}, \citenamefont {Pachitariu},
  \citenamefont {Burak}, \citenamefont {Baas}, \citenamefont {Dunn},
  \citenamefont {Moser},\ and\ \citenamefont {Moser}}]{gardner2022toroidal}%
  \BibitemOpen
  \bibfield  {author} {\bibinfo {author} {\bibfnamefont {R.~J.}\ \bibnamefont
  {Gardner}}, \bibinfo {author} {\bibfnamefont {E.}~\bibnamefont {Hermansen}},
  \bibinfo {author} {\bibfnamefont {M.}~\bibnamefont {Pachitariu}}, \bibinfo
  {author} {\bibfnamefont {Y.}~\bibnamefont {Burak}}, \bibinfo {author}
  {\bibfnamefont {N.~A.}\ \bibnamefont {Baas}}, \bibinfo {author}
  {\bibfnamefont {B.~A.}\ \bibnamefont {Dunn}}, \bibinfo {author}
  {\bibfnamefont {M.-B.}\ \bibnamefont {Moser}},\ and\ \bibinfo {author}
  {\bibfnamefont {E.~I.}\ \bibnamefont {Moser}},\ }\bibfield  {title} {\bibinfo
  {title} {Toroidal topology of population activity in grid cells},\
  }\href@noop {} {\bibfield  {journal} {\bibinfo  {journal} {Nature}\ }\textbf
  {\bibinfo {volume} {602}},\ \bibinfo {pages} {123} (\bibinfo {year}
  {2022})}\BibitemShut {NoStop}%
\bibitem [{\citenamefont {Dorrell}\ and\ \citenamefont
  {Whittington}(2026)}]{dorrell2026normative}%
  \BibitemOpen
  \bibfield  {author} {\bibinfo {author} {\bibfnamefont {W.}~\bibnamefont
  {Dorrell}}\ and\ \bibinfo {author} {\bibfnamefont {J.~C.}\ \bibnamefont
  {Whittington}},\ }\bibfield  {title} {\bibinfo {title} {If grid cells are the
  answer, what is the question? a review of normative grid cell theory},\
  }\href@noop {} {\bibfield  {journal} {\bibinfo  {journal} {arXiv preprint
  arXiv:2601.12424}\ } (\bibinfo {year} {2026})}\BibitemShut {NoStop}%
\bibitem [{\citenamefont {Moser}\ and\ \citenamefont
  {Moser}(2008)}]{moser2008metric}%
  \BibitemOpen
  \bibfield  {author} {\bibinfo {author} {\bibfnamefont {E.~I.}\ \bibnamefont
  {Moser}}\ and\ \bibinfo {author} {\bibfnamefont {M.-B.}\ \bibnamefont
  {Moser}},\ }\bibfield  {title} {\bibinfo {title} {A metric for space},\
  }\href@noop {} {\bibfield  {journal} {\bibinfo  {journal} {Hippocampus}\
  }\textbf {\bibinfo {volume} {18}},\ \bibinfo {pages} {1142} (\bibinfo {year}
  {2008})}\BibitemShut {NoStop}%
\bibitem [{\citenamefont {Ginosar}\ \emph {et~al.}(2023)\citenamefont
  {Ginosar}, \citenamefont {Aljadeff}, \citenamefont {Las}, \citenamefont
  {Derdikman},\ and\ \citenamefont {Ulanovsky}}]{ginosar2023grid}%
  \BibitemOpen
  \bibfield  {author} {\bibinfo {author} {\bibfnamefont {G.}~\bibnamefont
  {Ginosar}}, \bibinfo {author} {\bibfnamefont {J.}~\bibnamefont {Aljadeff}},
  \bibinfo {author} {\bibfnamefont {L.}~\bibnamefont {Las}}, \bibinfo {author}
  {\bibfnamefont {D.}~\bibnamefont {Derdikman}},\ and\ \bibinfo {author}
  {\bibfnamefont {N.}~\bibnamefont {Ulanovsky}},\ }\bibfield  {title} {\bibinfo
  {title} {Are grid cells used for navigation? on local metrics, subjective
  spaces, and black holes},\ }\href@noop {} {\bibfield  {journal} {\bibinfo
  {journal} {Neuron}\ }\textbf {\bibinfo {volume} {111}},\ \bibinfo {pages}
  {1858} (\bibinfo {year} {2023})}\BibitemShut {NoStop}%
\bibitem [{\citenamefont {Pettersen}\ \emph {et~al.}(2024)\citenamefont
  {Pettersen}, \citenamefont {Sch{\o}yen}, \citenamefont {{\O}stby},
  \citenamefont {Malthe-S{\o}renssen},\ and\ \citenamefont
  {Lepper{\o}d}}]{pettersen2024self}%
  \BibitemOpen
  \bibfield  {author} {\bibinfo {author} {\bibfnamefont {M.}~\bibnamefont
  {Pettersen}}, \bibinfo {author} {\bibfnamefont {V.~S.}\ \bibnamefont
  {Sch{\o}yen}}, \bibinfo {author} {\bibfnamefont {M.~D.}\ \bibnamefont
  {{\O}stby}}, \bibinfo {author} {\bibfnamefont {A.}~\bibnamefont
  {Malthe-S{\o}renssen}},\ and\ \bibinfo {author} {\bibfnamefont {M.~E.}\
  \bibnamefont {Lepper{\o}d}},\ }\bibfield  {title} {\bibinfo {title}
  {Self-supervised grid cells without path integration},\ }\href@noop {}
  {\bibfield  {journal} {\bibinfo  {journal} {bioRxiv}\ ,\ \bibinfo {pages}
  {2024}} (\bibinfo {year} {2024})}\BibitemShut {NoStop}%
\bibitem [{\citenamefont {Burak}\ and\ \citenamefont
  {Fiete}(2009)}]{burak2009accurate}%
  \BibitemOpen
  \bibfield  {author} {\bibinfo {author} {\bibfnamefont {Y.}~\bibnamefont
  {Burak}}\ and\ \bibinfo {author} {\bibfnamefont {I.~R.}\ \bibnamefont
  {Fiete}},\ }\bibfield  {title} {\bibinfo {title} {Accurate path integration
  in continuous attractor network models of grid cells},\ }\href@noop {}
  {\bibfield  {journal} {\bibinfo  {journal} {PLoS computational biology}\
  }\textbf {\bibinfo {volume} {5}},\ \bibinfo {pages} {e1000291} (\bibinfo
  {year} {2009})}\BibitemShut {NoStop}%
\bibitem [{\citenamefont {Stachenfeld}\ \emph {et~al.}(2017)\citenamefont
  {Stachenfeld}, \citenamefont {Botvinick},\ and\ \citenamefont
  {Gershman}}]{stachenfeld2017hippocampus}%
  \BibitemOpen
  \bibfield  {author} {\bibinfo {author} {\bibfnamefont {K.~L.}\ \bibnamefont
  {Stachenfeld}}, \bibinfo {author} {\bibfnamefont {M.~M.}\ \bibnamefont
  {Botvinick}},\ and\ \bibinfo {author} {\bibfnamefont {S.~J.}\ \bibnamefont
  {Gershman}},\ }\bibfield  {title} {\bibinfo {title} {The hippocampus as a
  predictive map},\ }\href@noop {} {\bibfield  {journal} {\bibinfo  {journal}
  {Nature neuroscience}\ }\textbf {\bibinfo {volume} {20}},\ \bibinfo {pages}
  {1643} (\bibinfo {year} {2017})}\BibitemShut {NoStop}%
\bibitem [{\citenamefont {Banino}\ \emph {et~al.}(2018)\citenamefont {Banino},
  \citenamefont {Barry}, \citenamefont {Uria}, \citenamefont {Blundell},
  \citenamefont {Lillicrap}, \citenamefont {Mirowski}, \citenamefont {Pritzel},
  \citenamefont {Chadwick}, \citenamefont {Degris}, \citenamefont {Modayil}
  \emph {et~al.}}]{banino2018vector}%
  \BibitemOpen
  \bibfield  {author} {\bibinfo {author} {\bibfnamefont {A.}~\bibnamefont
  {Banino}}, \bibinfo {author} {\bibfnamefont {C.}~\bibnamefont {Barry}},
  \bibinfo {author} {\bibfnamefont {B.}~\bibnamefont {Uria}}, \bibinfo {author}
  {\bibfnamefont {C.}~\bibnamefont {Blundell}}, \bibinfo {author}
  {\bibfnamefont {T.}~\bibnamefont {Lillicrap}}, \bibinfo {author}
  {\bibfnamefont {P.}~\bibnamefont {Mirowski}}, \bibinfo {author}
  {\bibfnamefont {A.}~\bibnamefont {Pritzel}}, \bibinfo {author} {\bibfnamefont
  {M.~J.}\ \bibnamefont {Chadwick}}, \bibinfo {author} {\bibfnamefont
  {T.}~\bibnamefont {Degris}}, \bibinfo {author} {\bibfnamefont
  {J.}~\bibnamefont {Modayil}}, \emph {et~al.},\ }\bibfield  {title} {\bibinfo
  {title} {Vector-based navigation using grid-like representations in
  artificial agents},\ }\href@noop {} {\bibfield  {journal} {\bibinfo
  {journal} {Nature}\ }\textbf {\bibinfo {volume} {557}},\ \bibinfo {pages}
  {429} (\bibinfo {year} {2018})}\BibitemShut {NoStop}%
\bibitem [{\citenamefont {Cueva}\ and\ \citenamefont
  {Wei}(2018)}]{cueva2018emergence}%
  \BibitemOpen
  \bibfield  {author} {\bibinfo {author} {\bibfnamefont {C.~J.}\ \bibnamefont
  {Cueva}}\ and\ \bibinfo {author} {\bibfnamefont {X.-X.}\ \bibnamefont
  {Wei}},\ }\bibfield  {title} {\bibinfo {title} {Emergence of grid-like
  representations by training recurrent neural networks to perform spatial
  localization},\ }\href@noop {} {\bibfield  {journal} {\bibinfo  {journal}
  {arXiv preprint arXiv:1803.07770}\ } (\bibinfo {year} {2018})}\BibitemShut
  {NoStop}%
\bibitem [{\citenamefont {Sorscher}\ \emph {et~al.}(2019)\citenamefont
  {Sorscher}, \citenamefont {Mel}, \citenamefont {Ganguli},\ and\ \citenamefont
  {Ocko}}]{sorscher2019unified}%
  \BibitemOpen
  \bibfield  {author} {\bibinfo {author} {\bibfnamefont {B.}~\bibnamefont
  {Sorscher}}, \bibinfo {author} {\bibfnamefont {G.}~\bibnamefont {Mel}},
  \bibinfo {author} {\bibfnamefont {S.}~\bibnamefont {Ganguli}},\ and\ \bibinfo
  {author} {\bibfnamefont {S.}~\bibnamefont {Ocko}},\ }\bibfield  {title}
  {\bibinfo {title} {A unified theory for the origin of grid cells through the
  lens of pattern formation},\ }\href@noop {} {\bibfield  {journal} {\bibinfo
  {journal} {Advances in neural information processing systems}\ }\textbf
  {\bibinfo {volume} {32}} (\bibinfo {year} {2019})}\BibitemShut {NoStop}%
\bibitem [{\citenamefont {Whittington}\ \emph {et~al.}(2020)\citenamefont
  {Whittington}, \citenamefont {Muller}, \citenamefont {Mark}, \citenamefont
  {Chen}, \citenamefont {Barry}, \citenamefont {Burgess},\ and\ \citenamefont
  {Behrens}}]{whittington2020tolman}%
  \BibitemOpen
  \bibfield  {author} {\bibinfo {author} {\bibfnamefont {J.~C.}\ \bibnamefont
  {Whittington}}, \bibinfo {author} {\bibfnamefont {T.~H.}\ \bibnamefont
  {Muller}}, \bibinfo {author} {\bibfnamefont {S.}~\bibnamefont {Mark}},
  \bibinfo {author} {\bibfnamefont {G.}~\bibnamefont {Chen}}, \bibinfo {author}
  {\bibfnamefont {C.}~\bibnamefont {Barry}}, \bibinfo {author} {\bibfnamefont
  {N.}~\bibnamefont {Burgess}},\ and\ \bibinfo {author} {\bibfnamefont {T.~E.}\
  \bibnamefont {Behrens}},\ }\bibfield  {title} {\bibinfo {title} {The
  tolman-eichenbaum machine: unifying space and relational memory through
  generalization in the hippocampal formation},\ }\href@noop {} {\bibfield
  {journal} {\bibinfo  {journal} {Cell}\ }\textbf {\bibinfo {volume} {183}},\
  \bibinfo {pages} {1249} (\bibinfo {year} {2020})}\BibitemShut {NoStop}%
\bibitem [{\citenamefont {Gao}\ \emph {et~al.}(2021)\citenamefont {Gao},
  \citenamefont {Xie}, \citenamefont {Wei}, \citenamefont {Zhu},\ and\
  \citenamefont {Wu}}]{gao2021path}%
  \BibitemOpen
  \bibfield  {author} {\bibinfo {author} {\bibfnamefont {R.}~\bibnamefont
  {Gao}}, \bibinfo {author} {\bibfnamefont {J.}~\bibnamefont {Xie}}, \bibinfo
  {author} {\bibfnamefont {X.-X.}\ \bibnamefont {Wei}}, \bibinfo {author}
  {\bibfnamefont {S.-C.}\ \bibnamefont {Zhu}},\ and\ \bibinfo {author}
  {\bibfnamefont {Y.~N.}\ \bibnamefont {Wu}},\ }\bibfield  {title} {\bibinfo
  {title} {On path integration of grid cells: group representation and
  isotropic scaling},\ }\href@noop {} {\bibfield  {journal} {\bibinfo
  {journal} {Advances in Neural Information Processing Systems}\ }\textbf
  {\bibinfo {volume} {34}},\ \bibinfo {pages} {28623} (\bibinfo {year}
  {2021})}\BibitemShut {NoStop}%
\bibitem [{\citenamefont {Xu}\ \emph {et~al.}(2023)\citenamefont {Xu},
  \citenamefont {Gao}, \citenamefont {Zhang}, \citenamefont {Wei},\ and\
  \citenamefont {Wu}}]{xu2023conformal}%
  \BibitemOpen
  \bibfield  {author} {\bibinfo {author} {\bibfnamefont {D.}~\bibnamefont
  {Xu}}, \bibinfo {author} {\bibfnamefont {R.}~\bibnamefont {Gao}}, \bibinfo
  {author} {\bibfnamefont {W.-H.}\ \bibnamefont {Zhang}}, \bibinfo {author}
  {\bibfnamefont {X.-X.}\ \bibnamefont {Wei}},\ and\ \bibinfo {author}
  {\bibfnamefont {Y.~N.}\ \bibnamefont {Wu}},\ }\bibfield  {title} {\bibinfo
  {title} {Conformal isometry of lie group representation in recurrent network
  of grid cells},\ }in\ \href@noop {} {\emph {\bibinfo {booktitle} {NeurIPS
  Workshop on Symmetry and Geometry in Neural Representations}}}\ (\bibinfo
  {organization} {PMLR},\ \bibinfo {year} {2023})\ pp.\ \bibinfo {pages}
  {370--387}\BibitemShut {NoStop}%
\bibitem [{\citenamefont {Dorrell}\ \emph {et~al.}(2022)\citenamefont
  {Dorrell}, \citenamefont {Latham}, \citenamefont {Behrens},\ and\
  \citenamefont {Whittington}}]{dorrell2022actionable}%
  \BibitemOpen
  \bibfield  {author} {\bibinfo {author} {\bibfnamefont {W.}~\bibnamefont
  {Dorrell}}, \bibinfo {author} {\bibfnamefont {P.~E.}\ \bibnamefont {Latham}},
  \bibinfo {author} {\bibfnamefont {T.~E.}\ \bibnamefont {Behrens}},\ and\
  \bibinfo {author} {\bibfnamefont {J.~C.}\ \bibnamefont {Whittington}},\
  }\bibfield  {title} {\bibinfo {title} {Actionable neural representations:
  Grid cells from minimal constraints},\ }\href@noop {} {\bibfield  {journal}
  {\bibinfo  {journal} {arXiv preprint arXiv:2209.15563}\ } (\bibinfo {year}
  {2022})}\BibitemShut {NoStop}%
\bibitem [{\citenamefont {Xu}\ \emph {et~al.}(2025)\citenamefont {Xu},
  \citenamefont {Gao}, \citenamefont {Zhang}, \citenamefont {Wei},\ and\
  \citenamefont {Wu}}]{xu2025conformal}%
  \BibitemOpen
  \bibfield  {author} {\bibinfo {author} {\bibfnamefont {D.}~\bibnamefont
  {Xu}}, \bibinfo {author} {\bibfnamefont {R.}~\bibnamefont {Gao}}, \bibinfo
  {author} {\bibfnamefont {W.}~\bibnamefont {Zhang}}, \bibinfo {author}
  {\bibfnamefont {X.-X.}\ \bibnamefont {Wei}},\ and\ \bibinfo {author}
  {\bibfnamefont {Y.}~\bibnamefont {Wu}},\ }\bibfield  {title} {\bibinfo
  {title} {On conformal isometry of grid cells: Learning distance-preserving
  position embedding},\ }in\ \href@noop {} {\emph {\bibinfo {booktitle}
  {International Conference on Learning Representations}}},\ Vol.\ \bibinfo
  {volume} {2025}\ (\bibinfo {year} {2025})\ pp.\ \bibinfo {pages}
  {30739--30761}\BibitemShut {NoStop}%
\bibitem [{\citenamefont {Schaeffer}\ \emph {et~al.}(2022)\citenamefont
  {Schaeffer}, \citenamefont {Khona},\ and\ \citenamefont
  {Fiete}}]{schaeffer2022no}%
  \BibitemOpen
  \bibfield  {author} {\bibinfo {author} {\bibfnamefont {R.}~\bibnamefont
  {Schaeffer}}, \bibinfo {author} {\bibfnamefont {M.}~\bibnamefont {Khona}},\
  and\ \bibinfo {author} {\bibfnamefont {I.}~\bibnamefont {Fiete}},\ }\bibfield
   {title} {\bibinfo {title} {No free lunch from deep learning in neuroscience:
  A case study through models of the entorhinal-hippocampal circuit},\
  }\href@noop {} {\bibfield  {journal} {\bibinfo  {journal} {Advances in neural
  information processing systems}\ }\textbf {\bibinfo {volume} {35}},\ \bibinfo
  {pages} {16052} (\bibinfo {year} {2022})}\BibitemShut {NoStop}%
\bibitem [{\citenamefont {Sch{\o}yen}\ \emph {et~al.}(2023)\citenamefont
  {Sch{\o}yen}, \citenamefont {Pettersen}, \citenamefont {Holzhausen},
  \citenamefont {Fyhn}, \citenamefont {Malthe-S{\o}renssen},\ and\
  \citenamefont {Lepper{\o}d}}]{schoyen2023coherently}%
  \BibitemOpen
  \bibfield  {author} {\bibinfo {author} {\bibfnamefont {V.}~\bibnamefont
  {Sch{\o}yen}}, \bibinfo {author} {\bibfnamefont {M.~B.}\ \bibnamefont
  {Pettersen}}, \bibinfo {author} {\bibfnamefont {K.}~\bibnamefont
  {Holzhausen}}, \bibinfo {author} {\bibfnamefont {M.}~\bibnamefont {Fyhn}},
  \bibinfo {author} {\bibfnamefont {A.}~\bibnamefont {Malthe-S{\o}renssen}},\
  and\ \bibinfo {author} {\bibfnamefont {M.~E.}\ \bibnamefont {Lepper{\o}d}},\
  }\bibfield  {title} {\bibinfo {title} {Coherently remapping toroidal cells
  but not grid cells are responsible for path integration in virtual agents},\
  }\href@noop {} {\bibfield  {journal} {\bibinfo  {journal} {Iscience}\
  }\textbf {\bibinfo {volume} {26}} (\bibinfo {year} {2023})}\BibitemShut
  {NoStop}%
\bibitem [{\citenamefont {Sorscher}\ \emph {et~al.}(2023)\citenamefont
  {Sorscher}, \citenamefont {Mel}, \citenamefont {Ocko}, \citenamefont
  {Giocomo},\ and\ \citenamefont {Ganguli}}]{sorscher2023unified}%
  \BibitemOpen
  \bibfield  {author} {\bibinfo {author} {\bibfnamefont {B.}~\bibnamefont
  {Sorscher}}, \bibinfo {author} {\bibfnamefont {G.~C.}\ \bibnamefont {Mel}},
  \bibinfo {author} {\bibfnamefont {S.~A.}\ \bibnamefont {Ocko}}, \bibinfo
  {author} {\bibfnamefont {L.~M.}\ \bibnamefont {Giocomo}},\ and\ \bibinfo
  {author} {\bibfnamefont {S.}~\bibnamefont {Ganguli}},\ }\bibfield  {title}
  {\bibinfo {title} {A unified theory for the computational and mechanistic
  origins of grid cells},\ }\href@noop {} {\bibfield  {journal} {\bibinfo
  {journal} {Neuron}\ }\textbf {\bibinfo {volume} {111}},\ \bibinfo {pages}
  {121} (\bibinfo {year} {2023})}\BibitemShut {NoStop}%
\bibitem [{\citenamefont {Schaeffer}\ \emph {et~al.}(2023)\citenamefont
  {Schaeffer}, \citenamefont {Khona}, \citenamefont {Ma}, \citenamefont
  {Eyzaguirre}, \citenamefont {Koyejo},\ and\ \citenamefont
  {Fiete}}]{schaeffer2023self}%
  \BibitemOpen
  \bibfield  {author} {\bibinfo {author} {\bibfnamefont {R.}~\bibnamefont
  {Schaeffer}}, \bibinfo {author} {\bibfnamefont {M.}~\bibnamefont {Khona}},
  \bibinfo {author} {\bibfnamefont {T.}~\bibnamefont {Ma}}, \bibinfo {author}
  {\bibfnamefont {C.}~\bibnamefont {Eyzaguirre}}, \bibinfo {author}
  {\bibfnamefont {S.}~\bibnamefont {Koyejo}},\ and\ \bibinfo {author}
  {\bibfnamefont {I.}~\bibnamefont {Fiete}},\ }\bibfield  {title} {\bibinfo
  {title} {Self-supervised learning of representations for space generates
  multi-modular grid cells},\ }\href@noop {} {\bibfield  {journal} {\bibinfo
  {journal} {Advances in Neural Information Processing Systems}\ }\textbf
  {\bibinfo {volume} {36}},\ \bibinfo {pages} {23140} (\bibinfo {year}
  {2023})}\BibitemShut {NoStop}%
\bibitem [{\citenamefont {Attneave}(1954)}]{attneave1954some}%
  \BibitemOpen
  \bibfield  {author} {\bibinfo {author} {\bibfnamefont {F.}~\bibnamefont
  {Attneave}},\ }\bibfield  {title} {\bibinfo {title} {Some informational
  aspects of visual perception.},\ }\href@noop {} {\bibfield  {journal}
  {\bibinfo  {journal} {Psychological review}\ }\textbf {\bibinfo {volume}
  {61}},\ \bibinfo {pages} {183} (\bibinfo {year} {1954})}\BibitemShut
  {NoStop}%
\bibitem [{\citenamefont {Barlow}\ \emph {et~al.}(1961)\citenamefont {Barlow}
  \emph {et~al.}}]{barlow1961possible}%
  \BibitemOpen
  \bibfield  {author} {\bibinfo {author} {\bibfnamefont {H.~B.}\ \bibnamefont
  {Barlow}} \emph {et~al.},\ }\bibfield  {title} {\bibinfo {title} {Possible
  principles underlying the transformation of sensory messages},\ }\href@noop
  {} {\bibfield  {journal} {\bibinfo  {journal} {Sensory communication}\
  }\textbf {\bibinfo {volume} {1}},\ \bibinfo {pages} {217} (\bibinfo {year}
  {1961})}\BibitemShut {NoStop}%
\bibitem [{\citenamefont {Shepard}\ and\ \citenamefont
  {Chipman}(1970)}]{shepard1970second}%
  \BibitemOpen
  \bibfield  {author} {\bibinfo {author} {\bibfnamefont {R.~N.}\ \bibnamefont
  {Shepard}}\ and\ \bibinfo {author} {\bibfnamefont {S.}~\bibnamefont
  {Chipman}},\ }\bibfield  {title} {\bibinfo {title} {Second-order isomorphism
  of internal representations: Shapes of states},\ }\href@noop {} {\bibfield
  {journal} {\bibinfo  {journal} {Cognitive psychology}\ }\textbf {\bibinfo
  {volume} {1}},\ \bibinfo {pages} {1} (\bibinfo {year} {1970})}\BibitemShut
  {NoStop}%
\bibitem [{\citenamefont {Edelman}(1998)}]{edelman1998representation}%
  \BibitemOpen
  \bibfield  {author} {\bibinfo {author} {\bibfnamefont {S.}~\bibnamefont
  {Edelman}},\ }\bibfield  {title} {\bibinfo {title} {Representation is
  representation of similarities},\ }\href@noop {} {\bibfield  {journal}
  {\bibinfo  {journal} {Behavioral and brain sciences}\ }\textbf {\bibinfo
  {volume} {21}},\ \bibinfo {pages} {449} (\bibinfo {year} {1998})}\BibitemShut
  {NoStop}%
\bibitem [{\citenamefont {Wolpert}\ and\ \citenamefont
  {Macready}(1997)}]{wolpert1997no}%
  \BibitemOpen
  \bibfield  {author} {\bibinfo {author} {\bibfnamefont {D.~H.}\ \bibnamefont
  {Wolpert}}\ and\ \bibinfo {author} {\bibfnamefont {W.~G.}\ \bibnamefont
  {Macready}},\ }\bibfield  {title} {\bibinfo {title} {No free lunch theorems
  for optimization},\ }\href@noop {} {\bibfield  {journal} {\bibinfo  {journal}
  {IEEE transactions on evolutionary computation}\ }\textbf {\bibinfo {volume}
  {1}},\ \bibinfo {pages} {67} (\bibinfo {year} {1997})}\BibitemShut {NoStop}%
\bibitem [{\citenamefont {Bengio}\ \emph {et~al.}(2013)\citenamefont {Bengio},
  \citenamefont {Courville},\ and\ \citenamefont
  {Vincent}}]{bengio2013representation}%
  \BibitemOpen
  \bibfield  {author} {\bibinfo {author} {\bibfnamefont {Y.}~\bibnamefont
  {Bengio}}, \bibinfo {author} {\bibfnamefont {A.}~\bibnamefont {Courville}},\
  and\ \bibinfo {author} {\bibfnamefont {P.}~\bibnamefont {Vincent}},\
  }\bibfield  {title} {\bibinfo {title} {Representation learning: A review and
  new perspectives},\ }\href@noop {} {\bibfield  {journal} {\bibinfo  {journal}
  {IEEE transactions on pattern analysis and machine intelligence}\ }\textbf
  {\bibinfo {volume} {35}},\ \bibinfo {pages} {1798} (\bibinfo {year}
  {2013})}\BibitemShut {NoStop}%
\bibitem [{\citenamefont {Bronstein}\ \emph {et~al.}(2021)\citenamefont
  {Bronstein}, \citenamefont {Bruna}, \citenamefont {Cohen},\ and\
  \citenamefont {Veli{\v{c}}kovi{\'c}}}]{bronstein2021geometric}%
  \BibitemOpen
  \bibfield  {author} {\bibinfo {author} {\bibfnamefont {M.~M.}\ \bibnamefont
  {Bronstein}}, \bibinfo {author} {\bibfnamefont {J.}~\bibnamefont {Bruna}},
  \bibinfo {author} {\bibfnamefont {T.}~\bibnamefont {Cohen}},\ and\ \bibinfo
  {author} {\bibfnamefont {P.}~\bibnamefont {Veli{\v{c}}kovi{\'c}}},\
  }\bibfield  {title} {\bibinfo {title} {Geometric deep learning: Grids,
  groups, graphs, geodesics, and gauges},\ }\href@noop {} {\bibfield  {journal}
  {\bibinfo  {journal} {arXiv preprint arXiv:2104.13478}\ } (\bibinfo {year}
  {2021})}\BibitemShut {NoStop}%
\bibitem [{\citenamefont {Sucholutsky}\ \emph {et~al.}(2023)\citenamefont
  {Sucholutsky}, \citenamefont {Muttenthaler}, \citenamefont {Weller},
  \citenamefont {Peng}, \citenamefont {Bobu}, \citenamefont {Kim},
  \citenamefont {Love}, \citenamefont {Grant}, \citenamefont {Groen},
  \citenamefont {Achterberg} \emph {et~al.}}]{sucholutsky2023getting}%
  \BibitemOpen
  \bibfield  {author} {\bibinfo {author} {\bibfnamefont {I.}~\bibnamefont
  {Sucholutsky}}, \bibinfo {author} {\bibfnamefont {L.}~\bibnamefont
  {Muttenthaler}}, \bibinfo {author} {\bibfnamefont {A.}~\bibnamefont
  {Weller}}, \bibinfo {author} {\bibfnamefont {A.}~\bibnamefont {Peng}},
  \bibinfo {author} {\bibfnamefont {A.}~\bibnamefont {Bobu}}, \bibinfo {author}
  {\bibfnamefont {B.}~\bibnamefont {Kim}}, \bibinfo {author} {\bibfnamefont
  {B.~C.}\ \bibnamefont {Love}}, \bibinfo {author} {\bibfnamefont
  {E.}~\bibnamefont {Grant}}, \bibinfo {author} {\bibfnamefont
  {I.}~\bibnamefont {Groen}}, \bibinfo {author} {\bibfnamefont
  {J.}~\bibnamefont {Achterberg}}, \emph {et~al.},\ }\bibfield  {title}
  {\bibinfo {title} {Getting aligned on representational alignment},\
  }\href@noop {} {\bibfield  {journal} {\bibinfo  {journal} {arXiv preprint
  arXiv:2310.13018}\ } (\bibinfo {year} {2023})}\BibitemShut {NoStop}%
\bibitem [{\citenamefont {Fiete}\ \emph {et~al.}(2008)\citenamefont {Fiete},
  \citenamefont {Burak},\ and\ \citenamefont {Brookings}}]{fiete2008grid}%
  \BibitemOpen
  \bibfield  {author} {\bibinfo {author} {\bibfnamefont {I.~R.}\ \bibnamefont
  {Fiete}}, \bibinfo {author} {\bibfnamefont {Y.}~\bibnamefont {Burak}},\ and\
  \bibinfo {author} {\bibfnamefont {T.}~\bibnamefont {Brookings}},\ }\bibfield
  {title} {\bibinfo {title} {What grid cells convey about rat location},\
  }\href@noop {} {\bibfield  {journal} {\bibinfo  {journal} {Journal of
  Neuroscience}\ }\textbf {\bibinfo {volume} {28}},\ \bibinfo {pages} {6858}
  (\bibinfo {year} {2008})}\BibitemShut {NoStop}%
\bibitem [{\citenamefont {Needham}(2021)}]{needham2021visual}%
  \BibitemOpen
  \bibfield  {author} {\bibinfo {author} {\bibfnamefont {T.}~\bibnamefont
  {Needham}},\ }\href@noop {} {\emph {\bibinfo {title} {Visual differential
  geometry and forms: a mathematical drama in five acts}}}\ (\bibinfo
  {publisher} {Princeton University Press},\ \bibinfo {year}
  {2021})\BibitemShut {NoStop}%
\bibitem [{\citenamefont {Hatcher}(2002)}]{hatcher2002algebraic}%
  \BibitemOpen
  \bibfield  {author} {\bibinfo {author} {\bibfnamefont {A.}~\bibnamefont
  {Hatcher}},\ }\href@noop {} {\emph {\bibinfo {title} {Algebraic Topology}}}\
  (\bibinfo  {publisher} {Cambridge University Press},\ \bibinfo {year}
  {2002})\BibitemShut {NoStop}%
\bibitem [{\citenamefont {Dwyer}\ and\ \citenamefont
  {Wilkerson}(1998)}]{dwyer1998elementary}%
  \BibitemOpen
  \bibfield  {author} {\bibinfo {author} {\bibfnamefont {W.~G.}\ \bibnamefont
  {Dwyer}}\ and\ \bibinfo {author} {\bibfnamefont {C.~W.}\ \bibnamefont
  {Wilkerson}},\ }\bibfield  {title} {\bibinfo {title} {The elementary
  geometric structure of compact lie groups},\ }\href@noop {} {\bibfield
  {journal} {\bibinfo  {journal} {Bulletin of the London Mathematical Society}\
  }\textbf {\bibinfo {volume} {30}},\ \bibinfo {pages} {337} (\bibinfo {year}
  {1998})}\BibitemShut {NoStop}%
\bibitem [{\citenamefont {Rank}(1984)}]{rank1984head}%
  \BibitemOpen
  \bibfield  {author} {\bibinfo {author} {\bibfnamefont {J.}~\bibnamefont
  {Rank}},\ }\bibfield  {title} {\bibinfo {title} {Head-direction cells in the
  deep layers of dorsal presubiculum of freely moving rats},\ }in\ \href@noop
  {} {\emph {\bibinfo {booktitle} {Soc. Neuroscience Abstr.}}},\ Vol.~\bibinfo
  {volume} {10}\ (\bibinfo {year} {1984})\ p.\ \bibinfo {pages}
  {599}\BibitemShut {NoStop}%
\bibitem [{\citenamefont {Ginosar}\ \emph {et~al.}(2021)\citenamefont
  {Ginosar}, \citenamefont {Aljadeff}, \citenamefont {Burak}, \citenamefont
  {Sompolinsky}, \citenamefont {Las},\ and\ \citenamefont
  {Ulanovsky}}]{ginosar2021locally}%
  \BibitemOpen
  \bibfield  {author} {\bibinfo {author} {\bibfnamefont {G.}~\bibnamefont
  {Ginosar}}, \bibinfo {author} {\bibfnamefont {J.}~\bibnamefont {Aljadeff}},
  \bibinfo {author} {\bibfnamefont {Y.}~\bibnamefont {Burak}}, \bibinfo
  {author} {\bibfnamefont {H.}~\bibnamefont {Sompolinsky}}, \bibinfo {author}
  {\bibfnamefont {L.}~\bibnamefont {Las}},\ and\ \bibinfo {author}
  {\bibfnamefont {N.}~\bibnamefont {Ulanovsky}},\ }\bibfield  {title} {\bibinfo
  {title} {Locally ordered representation of 3d space in the entorhinal
  cortex},\ }\href@noop {} {\bibfield  {journal} {\bibinfo  {journal} {Nature}\
  }\textbf {\bibinfo {volume} {596}},\ \bibinfo {pages} {404} (\bibinfo {year}
  {2021})}\BibitemShut {NoStop}%
\bibitem [{\citenamefont {Grieves}\ \emph {et~al.}(2021)\citenamefont
  {Grieves}, \citenamefont {Jedidi-Ayoub}, \citenamefont {Mishchanchuk},
  \citenamefont {Liu}, \citenamefont {Renaudineau}, \citenamefont {Duvelle},\
  and\ \citenamefont {Jeffery}}]{grieves2021irregular}%
  \BibitemOpen
  \bibfield  {author} {\bibinfo {author} {\bibfnamefont {R.~M.}\ \bibnamefont
  {Grieves}}, \bibinfo {author} {\bibfnamefont {S.}~\bibnamefont
  {Jedidi-Ayoub}}, \bibinfo {author} {\bibfnamefont {K.}~\bibnamefont
  {Mishchanchuk}}, \bibinfo {author} {\bibfnamefont {A.}~\bibnamefont {Liu}},
  \bibinfo {author} {\bibfnamefont {S.}~\bibnamefont {Renaudineau}}, \bibinfo
  {author} {\bibfnamefont {{\'E}.}~\bibnamefont {Duvelle}},\ and\ \bibinfo
  {author} {\bibfnamefont {K.~J.}\ \bibnamefont {Jeffery}},\ }\bibfield
  {title} {\bibinfo {title} {Irregular distribution of grid cell firing fields
  in rats exploring a 3d volumetric space},\ }\href@noop {} {\bibfield
  {journal} {\bibinfo  {journal} {Nature neuroscience}\ }\textbf {\bibinfo
  {volume} {24}},\ \bibinfo {pages} {1567} (\bibinfo {year}
  {2021})}\BibitemShut {NoStop}%
\bibitem [{\citenamefont {Qi}\ and\ \citenamefont {Yartsev}(2026)}]{qi2026two}%
  \BibitemOpen
  \bibfield  {author} {\bibinfo {author} {\bibfnamefont {K.~K.}\ \bibnamefont
  {Qi}}\ and\ \bibinfo {author} {\bibfnamefont {M.~M.}\ \bibnamefont
  {Yartsev}},\ }\bibfield  {title} {\bibinfo {title} {A two-dimensional
  grid-cell code for three-dimensional navigation in freely flying bats},\
  }\href@noop {} {\bibfield  {journal} {\bibinfo  {journal} {bioRxiv}\ ,\
  \bibinfo {pages} {2026.06.10.728358}} (\bibinfo {year} {2026})},\ \bibinfo
  {note} {preprint}\BibitemShut {NoStop}%
\bibitem [{\citenamefont {Skaggs}\ \emph {et~al.}(1994)\citenamefont {Skaggs},
  \citenamefont {Knierim}, \citenamefont {Kudrimoti},\ and\ \citenamefont
  {McNaughton}}]{skaggs1994model}%
  \BibitemOpen
  \bibfield  {author} {\bibinfo {author} {\bibfnamefont {W.}~\bibnamefont
  {Skaggs}}, \bibinfo {author} {\bibfnamefont {J.}~\bibnamefont {Knierim}},
  \bibinfo {author} {\bibfnamefont {H.}~\bibnamefont {Kudrimoti}},\ and\
  \bibinfo {author} {\bibfnamefont {B.}~\bibnamefont {McNaughton}},\ }\bibfield
   {title} {\bibinfo {title} {A model of the neural basis of the rat's sense of
  direction},\ }\href@noop {} {\bibfield  {journal} {\bibinfo  {journal}
  {Advances in neural information processing systems}\ }\textbf {\bibinfo
  {volume} {7}} (\bibinfo {year} {1994})}\BibitemShut {NoStop}%
\bibitem [{\citenamefont {Zhang}(1996)}]{zhang1996representation}%
  \BibitemOpen
  \bibfield  {author} {\bibinfo {author} {\bibfnamefont {K.}~\bibnamefont
  {Zhang}},\ }\bibfield  {title} {\bibinfo {title} {Representation of spatial
  orientation by the intrinsic dynamics of the head-direction cell ensemble: a
  theory},\ }\href@noop {} {\bibfield  {journal} {\bibinfo  {journal} {Journal
  of Neuroscience}\ }\textbf {\bibinfo {volume} {16}},\ \bibinfo {pages} {2112}
  (\bibinfo {year} {1996})}\BibitemShut {NoStop}%
\bibitem [{\citenamefont {Krupic}\ \emph {et~al.}(2015)\citenamefont {Krupic},
  \citenamefont {Bauza}, \citenamefont {Burton}, \citenamefont {Barry},\ and\
  \citenamefont {O’Keefe}}]{krupic2015grid}%
  \BibitemOpen
  \bibfield  {author} {\bibinfo {author} {\bibfnamefont {J.}~\bibnamefont
  {Krupic}}, \bibinfo {author} {\bibfnamefont {M.}~\bibnamefont {Bauza}},
  \bibinfo {author} {\bibfnamefont {S.}~\bibnamefont {Burton}}, \bibinfo
  {author} {\bibfnamefont {C.}~\bibnamefont {Barry}},\ and\ \bibinfo {author}
  {\bibfnamefont {J.}~\bibnamefont {O’Keefe}},\ }\bibfield  {title} {\bibinfo
  {title} {Grid cell symmetry is shaped by environmental geometry},\
  }\href@noop {} {\bibfield  {journal} {\bibinfo  {journal} {Nature}\ }\textbf
  {\bibinfo {volume} {518}},\ \bibinfo {pages} {232} (\bibinfo {year}
  {2015})}\BibitemShut {NoStop}%
\bibitem [{\citenamefont {Krupic}\ \emph {et~al.}(2018)\citenamefont {Krupic},
  \citenamefont {Bauza}, \citenamefont {Burton},\ and\ \citenamefont
  {O’Keefe}}]{krupic2018local}%
  \BibitemOpen
  \bibfield  {author} {\bibinfo {author} {\bibfnamefont {J.}~\bibnamefont
  {Krupic}}, \bibinfo {author} {\bibfnamefont {M.}~\bibnamefont {Bauza}},
  \bibinfo {author} {\bibfnamefont {S.}~\bibnamefont {Burton}},\ and\ \bibinfo
  {author} {\bibfnamefont {J.}~\bibnamefont {O’Keefe}},\ }\bibfield  {title}
  {\bibinfo {title} {Local transformations of the hippocampal cognitive map},\
  }\href@noop {} {\bibfield  {journal} {\bibinfo  {journal} {Science}\ }\textbf
  {\bibinfo {volume} {359}},\ \bibinfo {pages} {1143} (\bibinfo {year}
  {2018})}\BibitemShut {NoStop}%
\bibitem [{\citenamefont {Stensola}\ \emph {et~al.}(2015)\citenamefont
  {Stensola}, \citenamefont {Stensola}, \citenamefont {Moser},\ and\
  \citenamefont {Moser}}]{stensola2015shearing}%
  \BibitemOpen
  \bibfield  {author} {\bibinfo {author} {\bibfnamefont {T.}~\bibnamefont
  {Stensola}}, \bibinfo {author} {\bibfnamefont {H.}~\bibnamefont {Stensola}},
  \bibinfo {author} {\bibfnamefont {M.-B.}\ \bibnamefont {Moser}},\ and\
  \bibinfo {author} {\bibfnamefont {E.~I.}\ \bibnamefont {Moser}},\ }\bibfield
  {title} {\bibinfo {title} {Shearing-induced asymmetry in entorhinal grid
  cells},\ }\href@noop {} {\bibfield  {journal} {\bibinfo  {journal} {Nature}\
  }\textbf {\bibinfo {volume} {518}},\ \bibinfo {pages} {207} (\bibinfo {year}
  {2015})}\BibitemShut {NoStop}%
\bibitem [{\citenamefont {Sreenivasan}\ and\ \citenamefont
  {Fiete}(2011)}]{sreenivasan2011grid}%
  \BibitemOpen
  \bibfield  {author} {\bibinfo {author} {\bibfnamefont {S.}~\bibnamefont
  {Sreenivasan}}\ and\ \bibinfo {author} {\bibfnamefont {I.}~\bibnamefont
  {Fiete}},\ }\bibfield  {title} {\bibinfo {title} {Grid cells generate an
  analog error-correcting code for singularly precise neural computation},\
  }\href@noop {} {\bibfield  {journal} {\bibinfo  {journal} {Nature
  neuroscience}\ }\textbf {\bibinfo {volume} {14}},\ \bibinfo {pages} {1330}
  (\bibinfo {year} {2011})}\BibitemShut {NoStop}%
\bibitem [{\citenamefont {Pouget}\ \emph {et~al.}(2000)\citenamefont {Pouget},
  \citenamefont {Dayan},\ and\ \citenamefont {Zemel}}]{pouget2000information}%
  \BibitemOpen
  \bibfield  {author} {\bibinfo {author} {\bibfnamefont {A.}~\bibnamefont
  {Pouget}}, \bibinfo {author} {\bibfnamefont {P.}~\bibnamefont {Dayan}},\ and\
  \bibinfo {author} {\bibfnamefont {R.}~\bibnamefont {Zemel}},\ }\bibfield
  {title} {\bibinfo {title} {Information processing with population codes},\
  }\href@noop {} {\bibfield  {journal} {\bibinfo  {journal} {Nature Reviews
  Neuroscience}\ }\textbf {\bibinfo {volume} {1}},\ \bibinfo {pages} {125}
  (\bibinfo {year} {2000})}\BibitemShut {NoStop}%
\end{thebibliography}%
\newpage

\appendix

\section{TopoCN Model Details}\label{sec:num_setup}

\textbf{Torus embedding.} Consider a square-shaped 2D arena. Each location $(x,y)$ is mapped to a 4D toroidal embedding $\mathbf{n} \in \mathbb{R}^4$,
\begin{align}\label{eq:torus_embed}
  n_1 &= R\cos(k_1 x + k_2 y), \quad n_2 = R\sin(k_1x + k_2 y),\nonumber\\
  n_3 &= r\cos(k_3x + k_4 y), \quad n_4 = r\sin(k_3 x + k_4 y),
\end{align}
where $R, r, k_i$ are learnable parameters: $R$ and $r$ set the major and minor radii of the torus, and the $k_i$ specify the linear map from physical coordinates to torus angles. This guarantees that the network input already lies on a $T^2$ manifold, from which downstream layers can perform diverse transformations.

\textbf{Torus size.} The torus size $s$ quantifies the compactness of the representation through the mean population-activity vector,
\begin{equation}\label{eq:torus_size}
  s = 1 - \left\| \frac{1}{m} \sum_{k=1}^m \mathbf{g}^{(k)} \right\|_2^2,
\end{equation}
where $\mathbf{g}^{(k)}$ is the output representation of the $k$-th randomly sampled location. Since $\|\mathbf{g}\|_2=1$, we have $s \in [0,1]$, with larger values indicating activity more broadly distributed over the unit sphere (a more spread-out torus).


\textbf{Environment and data generation.} 
We modeled the environment as a two-dimensional square plane. During training, batches of positions were uniformly sampled from $[-2\pi,2\pi]$; during evaluation, inputs were taken from a regular $64 \times 64$ meshgrid covering the same domain. 

\textbf{Network details.} 
The input $(x,y)$ coordinates were first encoded by a custom \texttt{ToriActivation} layer implementing the embedding according Eq.~\eqref{eq:torus_embed}, and then passed through a multilayer perceptron with ReLU activations. During training, Gaussian noise was added to the last layer as a regularizer (distinct from the evaluation-time noise sweep described below), and the final activity vector was normalized to enforce the $\ell_2$ constraint ($\|\mathbf{g}\|_2=1$); non-negativity followed from the ReLU nonlinearity. The network thus implements a minimal feedforward mapping from 2D coordinates to a $G$-dimensional population vector $\mathbf{g}$ ($G=32$ in the main text).

\textbf{Training procedure.} 
We trained the model using the Adam optimizer with learning rate $10^{-3}$, batch size $64$, and $50{,}000$ training epochs. 

\textbf{Grid analysis.} 
To quantify the learned representations, we computed grid score, orientation and spacing of unit rate maps following the procedure of \citet{pettersen2024self}. Rate maps ($64 \times 64$) were smoothed with a Gaussian kernel. Grid score was defined as the difference between autocorrelogram correlations at $60^\circ$ and $30^\circ$; orientation as the smallest angle between the horizontal and the six innermost peaks (excluding the origin); spacing as the average distance to these peaks; and phase as the displacement of the nearest peak from the origin.

\textbf{Noise resilience.}
To probe robustness to neural variability, at evaluation we added independent Gaussian noise to each component of the output population vector $\mathbf{g}$ and renormalized to restore the constraint $\|\mathbf{g}\|_2 = 1$ before recomputing rate maps. We swept the noise standard deviation over the range shown in Fig.~\ref{fig:grid_statis}a, and at each level recorded the fraction of active cells (units whose mean rate exceeds $0.5$) and the fraction of active cells classified as grid cells (grid score $>0.7$).

\textbf{Hardware.}
All the models were trained on NVIDIA GeForce RTX 4090 (24 GB). For a single
model, training time was less than 15 minutes.

\section{RNN-based Path Integration}
\label{sec:PI}

\begin{figure}[h]
  \centering
  \begin{subfigure}{0.48\linewidth}
    \centering
    \includegraphics[width=\linewidth]{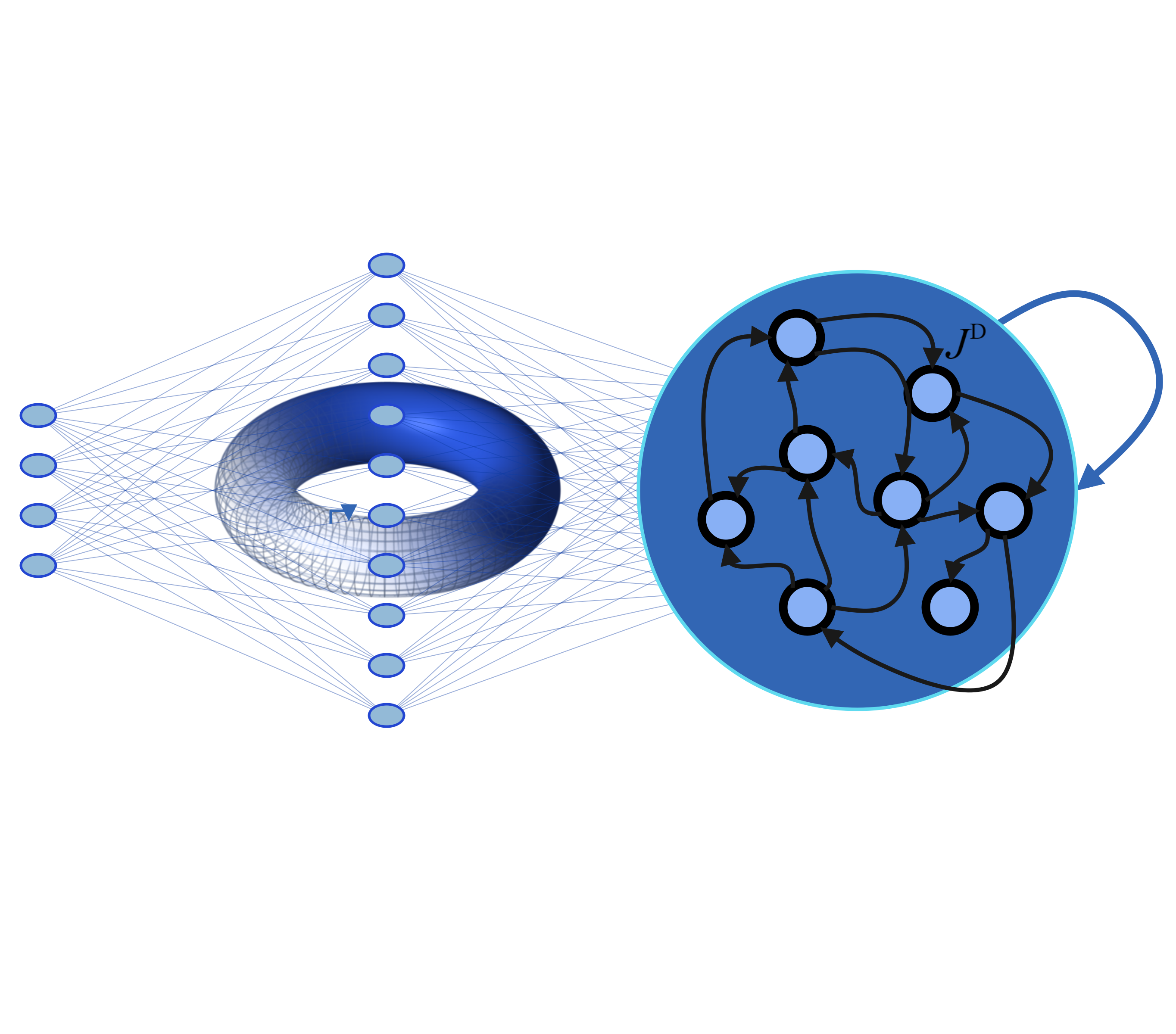}
    \caption{}
  \end{subfigure}
  \hfill
  \begin{subfigure}{0.48\linewidth}
    \centering
    \includegraphics[width=\linewidth]{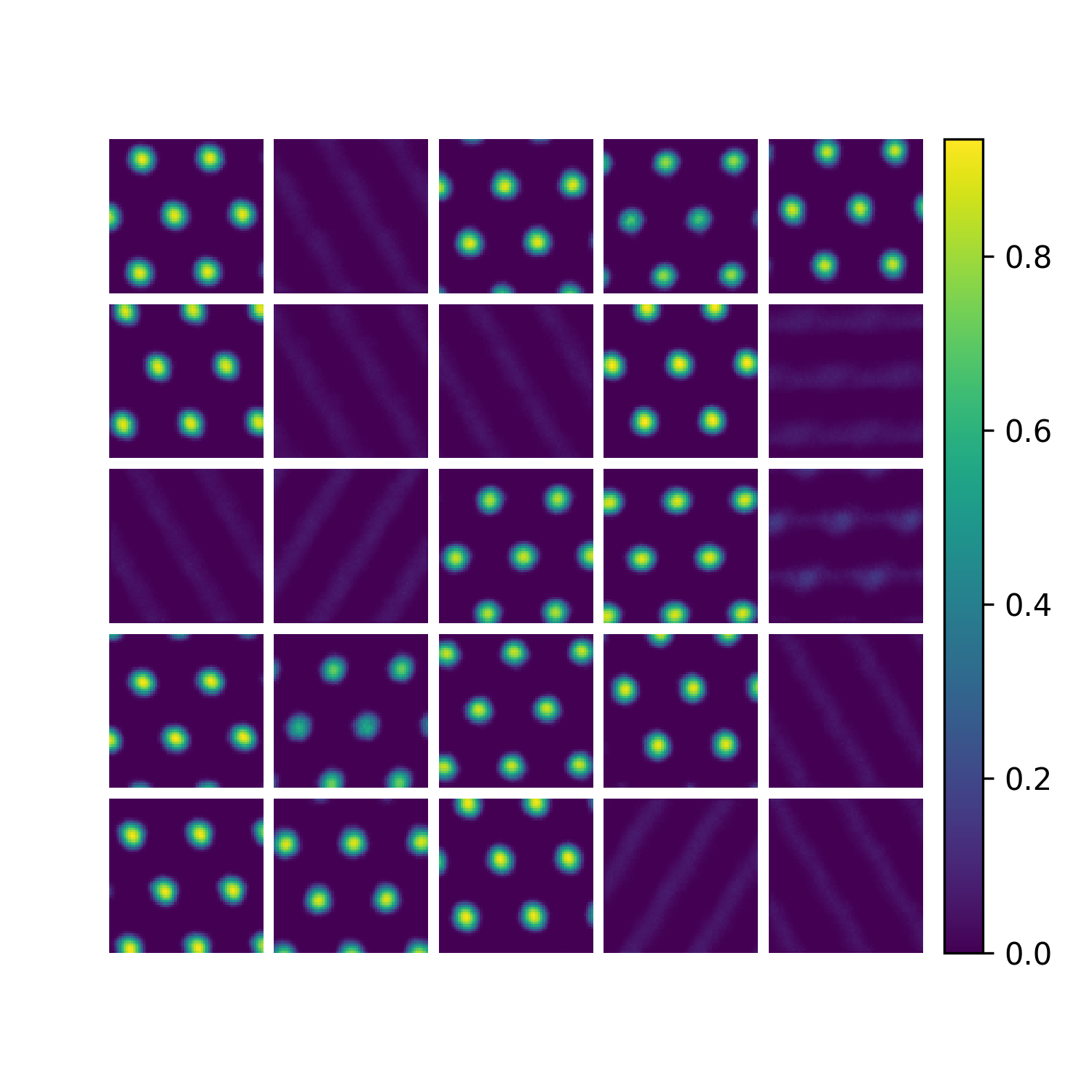}
    \label{fig:rnn_field}
    \caption{}
  \end{subfigure}
  \caption{RNN-based TopoCN and fring maps. (a) Architecture of the RNN-based TopoCN. (b) Example firing maps of individual neurons generated by the RNN-based TopoCN.}
\label{fig:rnn_combined}
\end{figure}


To assess the robustness of the emergence of hexagonal firing patterns, we extended the framework by coupling the TopoCN with a Recurrent Neural Network (RNN) designed for path integration \citep{pettersen2024self,xu2025conformal}. In this architecture, the TopoCN encodes the initial position on the toroidal manifold, while the RNN iteratively updates this state by integrating stepwise displacement inputs.

The model was trained using the same CI objective as in the main experiments (Eq.~\ref{eq:loss}). We observed that hexagonal firing patterns emerged robustly in this RNN regime, notably without the need for the explicit torus-size regularization or capacity penalties required in the static model (Fig.~\ref{fig:rnn_combined}). This suggests that the requirement for consistent temporal integration acts as an implicit geometric constraint. In the following subsection, we elucidate this phenomenon by analyzing how the interplay between recurrent dynamics and the CI constraint naturally drives the toroidal manifold toward a stable intrinsic size.

\section{Torus Size versus Capacity Regularization}\label{sec:cap}

\begin{figure}[h]
  \centering
  \begin{subfigure}{0.32\linewidth}
    \centering
    \includegraphics[width=\linewidth]{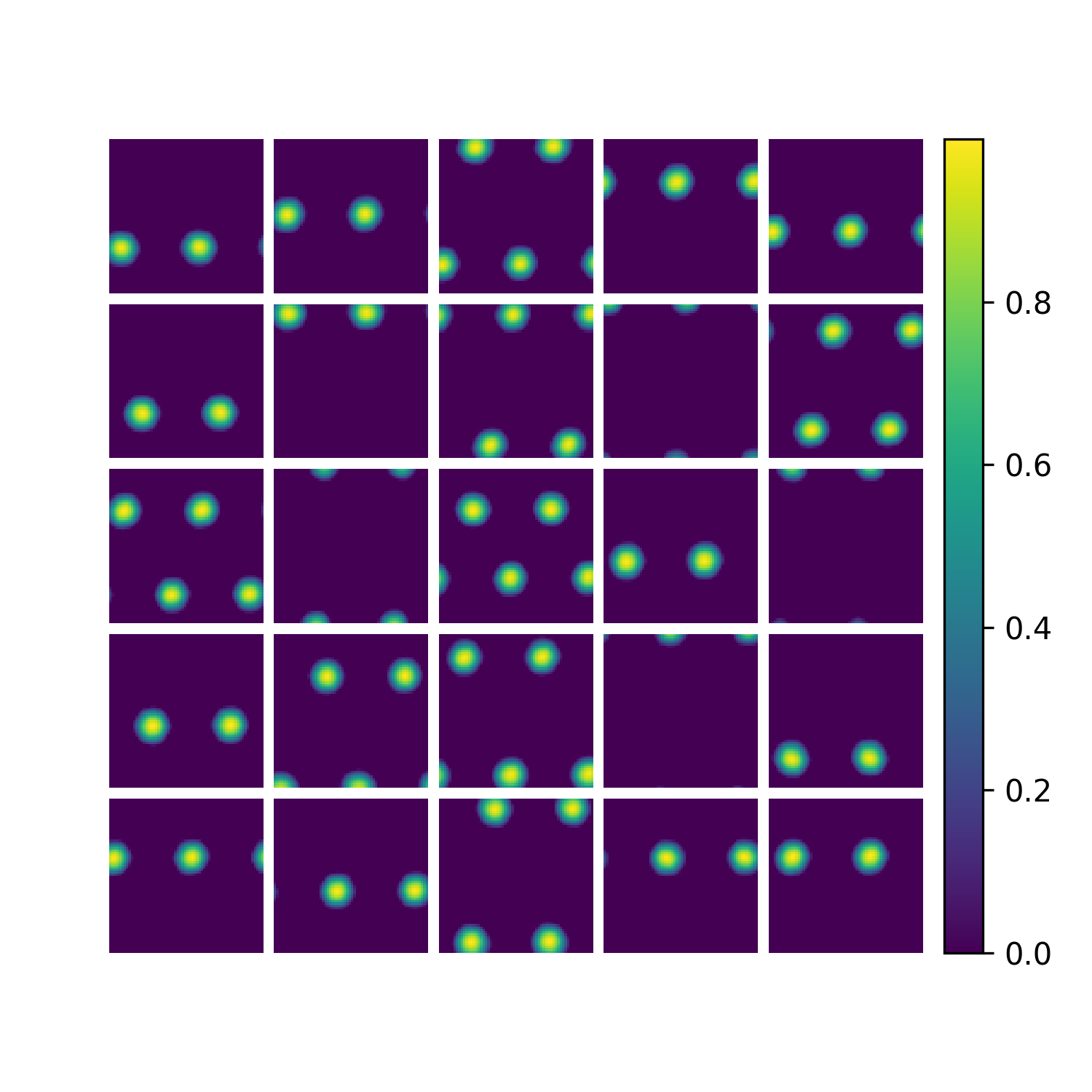}
    \caption{No regularization}
    \label{fig:ff_nosize}
  \end{subfigure}
  \hfill
  \begin{subfigure}{0.32\linewidth}
    \centering
    \includegraphics[width=\linewidth]{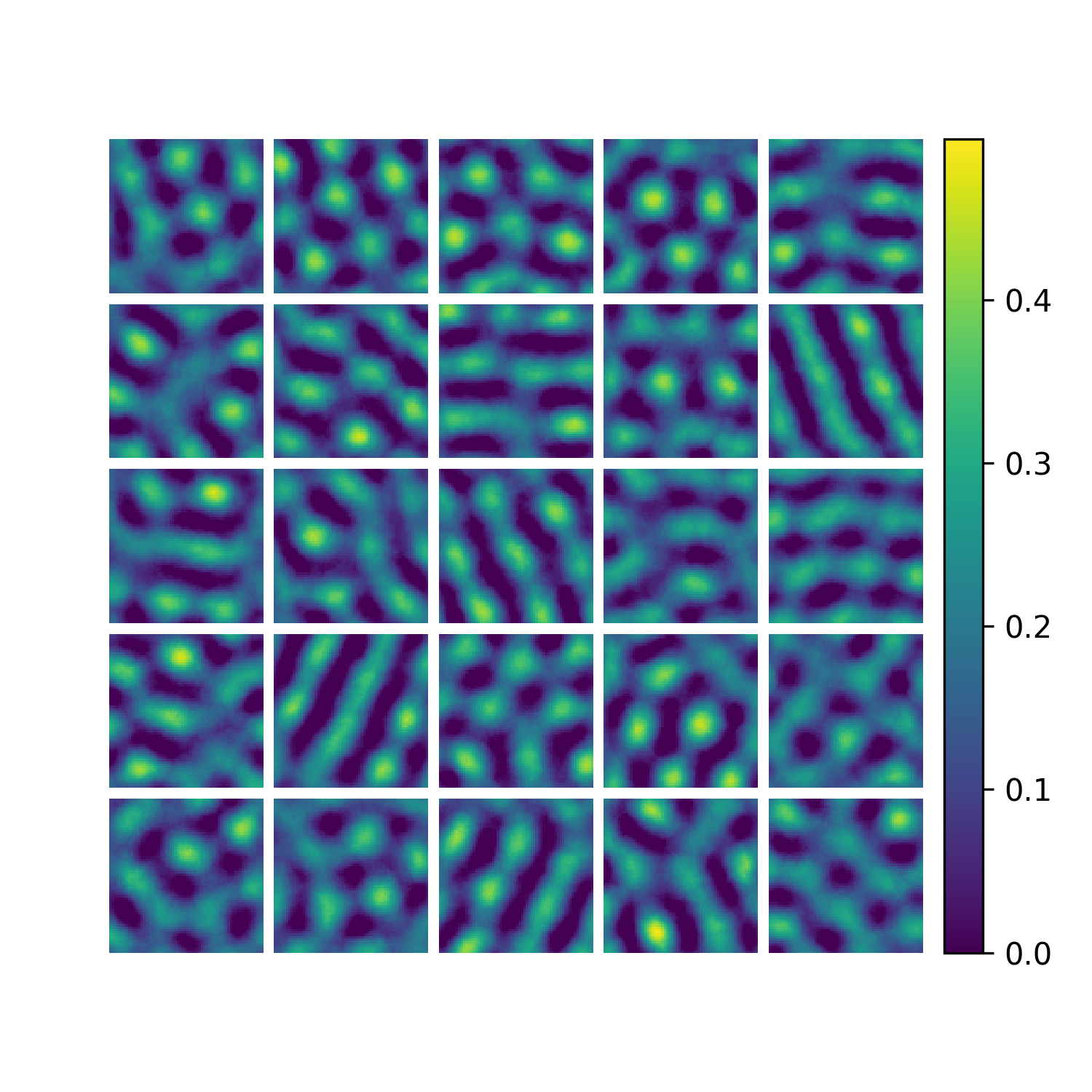}
    \caption{$L^1$ capacity regularization}
    \label{fig:ff_l1}
  \end{subfigure}
  \hfill
  \begin{subfigure}{0.32\linewidth}
    \centering
    \includegraphics[width=\linewidth]{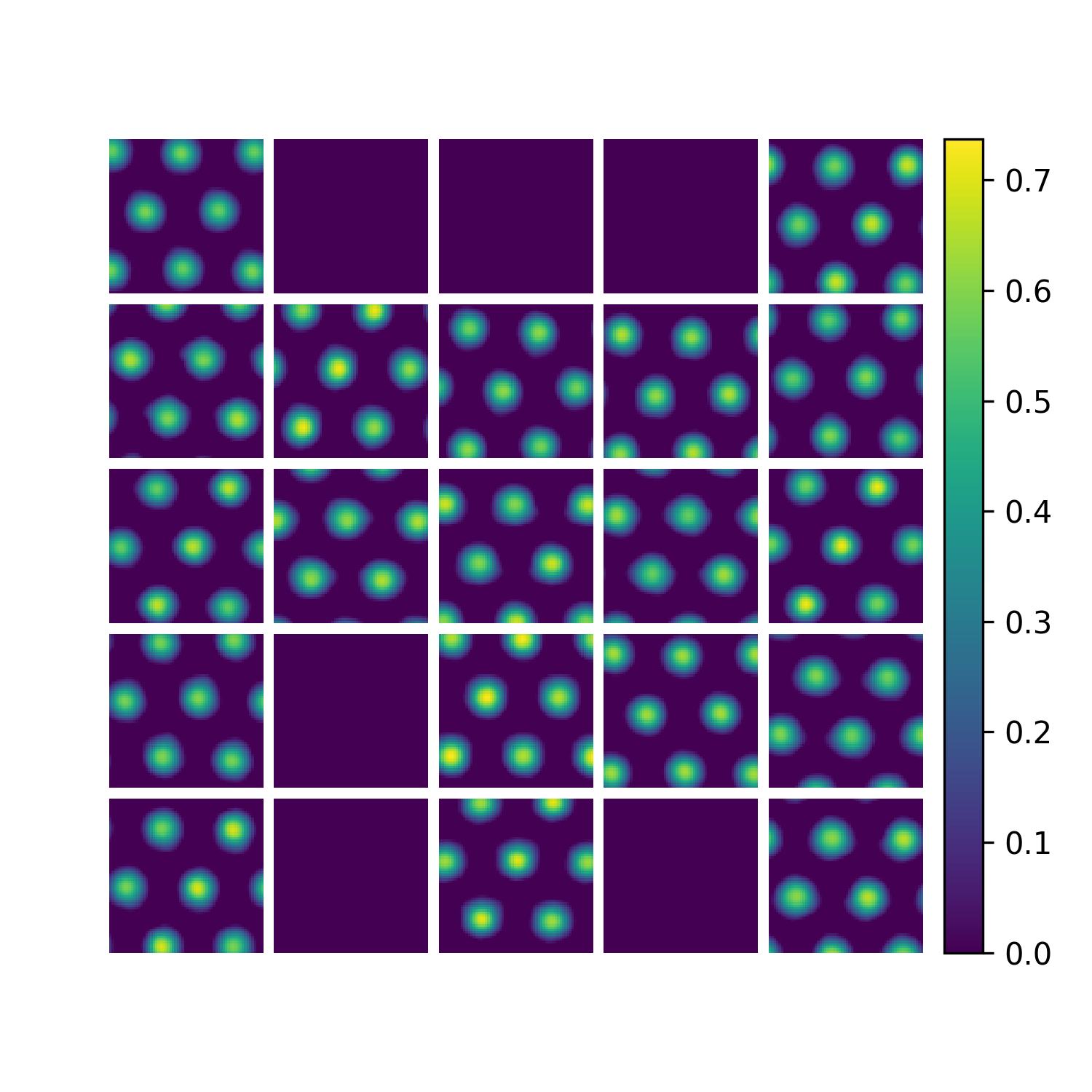}
    \caption{$L^2$ capacity regularization}
    \label{fig:ff_l2}
  \end{subfigure}
  \caption{Firing maps under different regularization schemes. Example firing maps produced by TopoCN trained with different regularization schemes.}
  \label{fig:ff_regs}
\end{figure}

We compared three training regimes on TopoCN: (i) no torus-size regularization; (ii) replacing the size term (the second term in \eqref{eq:loss}) with an $L^1$ capacity regularization\citep{pettersen2024self}; and (iii) replacing the size term with an $L^2$ capacity regularization \citep{schaeffer2023self}. The outcomes were consistent across simulations (Fig.~\ref{fig:ff_regs}):  

\begin{itemize}
    \item No regularization: The grid spacing became \emph{non-uniform}, and the firing patterns tended to be more \emph{square-like} than hexagonal;
    \item \textbf{$L^1$}: The firing patterns were generally more \emph{diffuse};
    \item \textbf{$L^2$}: The \emph{hexagonal} grid-like patterns emerged clearly.
\end{itemize}

Formally, the two types of capacity regularization can be written as follows. The $L^2$ version \citep{schaeffer2023self} penalizes the squared norm of the mean population activity:
\begin{equation}\label{eq:cap_L2}
    \mathcal{L}_{\text{cap}}^{(2)} \;=\; - \left\| \frac{1}{n} \sum_{k=1}^n \mathbf{g}^{(k)} \right\|_2^2,
\end{equation}
which encourages broad, uniform representations on the hypersphere and is closely related to our torus size measure in Eq.~\eqref{eq:torus_size}.  

The $L^1$ capacity constraint promotes maximally correlated activity across neurons, effectively pushing the population activity toward the diagonal of the state space where all units are coactive:
\begin{equation}\label{eq:cap_L1}
    \mathcal{L}_{\text{cap}}^{(1)} \;=\; - \sum_i g_i,
\end{equation}

These results have a common explanation in terms of torus size. The $L^2$ penalty in Eq.~\eqref{eq:cap_L2} equals the torus size $s$ measured in Eq.~\eqref{eq:torus_size} up to an additive constant, $s = 1 + \mathcal{L}_{\text{cap}}^{(2)}$, so adding it to the objective directly controls $s$; at the penalty strength used here it places $s$ in the intermediate regime where hexagonal fields emerge. The RNN experiments behave the same way: even \emph{without} an explicit penalty, the interplay between the CI loss and recurrent dynamics self-organizes $s$ into the same regime. In both cases the capacity term influences hexagonality \emph{through} torus size, and whether a given penalty yields hexagonal fields depends on whether its strength places $s$ in this window.

This motivates the choice made in the main text. Rather than tuning a capacity penalty and reading off whichever torus size it happens to produce, we treat $s_0$ as an explicit, interpretable control parameter and sweep it systematically, making the dependence of single-cell tuning on torus geometry transparent and reproducible.

\section{Practical handling of \texorpdfstring{$\rho$}{rho}} \label{sec:prac}

As shown in the main text, grid spacing decreases with the scale factor $\rho$ (roughly as $1/\rho$) when torus size $s_0$ is fixed, and increases with $s_0$ when $\rho$ is fixed (Fig.~\ref{fig:spacing}).  

In practice, however, the loss function in Eq.~\eqref{eq:loss} uses chord length in neural space rather than true geodesic length. This introduces a small mismatch: the actual scale realized by the trained network, denoted $\rho_{\text{true}}$, may differ from the nominal $\rho$ set during training. As a result, spacing measured from neural activity is determined by $\rho_{\text{true}}$, not the nominal value.  

\begin{figure}[h]
  \centering
  \includegraphics[width=\linewidth]{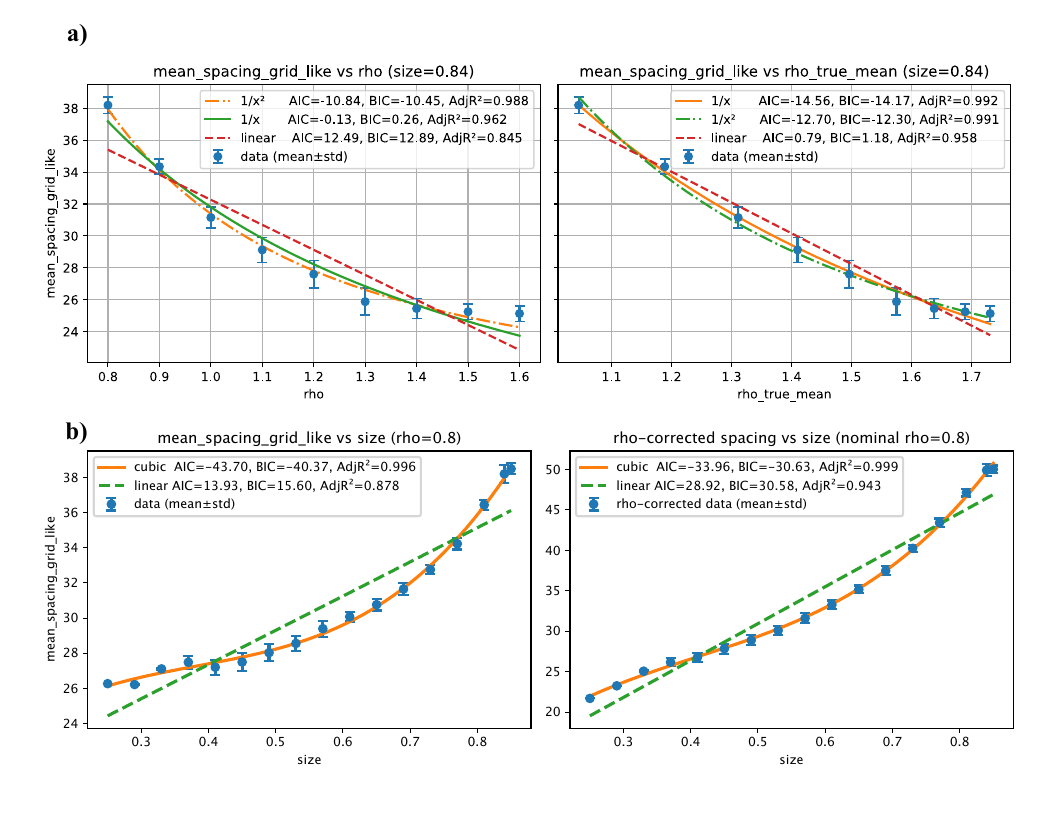}
  \caption{Effective Mapping Scale Explains the Spacing--Size Relationship. (a) Comparison of grid spacing as a function of nominal $\rho$ and effective $\rho$. (b) Relationship between grid spacing and torus size before and after correcting for effective $\rho$.}
  \label{fig:spacing_compare}
\end{figure}

To deal with this, we distinguish two cases:

\textbf{Varying $\rho$ with fixed $s_0$:}  
We compare spacing as a function of the \emph{nominal} $\rho$ and the \emph{measured} $\rho_{\text{true}}$, and fit the latter with the predicted hyperbolic law $\,\text{sp} \propto 1/\rho$. This shows that the effective scale realized by the network follows the theoretical prediction (Fig.~\ref{fig:spacing_compare}a).

\textbf{Varying $s_0$ with fixed $\rho$:}  
Even if $\rho$ is fixed nominally, small deviations in $\rho_{\text{true}}$ can still affect spacing. To remove this confound, we normalize spacing to a common reference scale $\rho_{\text{ref}}$ using
\begin{equation}
\label{eq:rho_correction}
\text{sp}_{\text{corr}} = \text{sp} \times \frac{\rho_{\text{true}}}{\rho_{\text{ref}}},
\end{equation}
which follows from the proportionality $\text{sp} \propto 1/\rho$.  
We therefore report both the raw spacing–vs–size curve and the $\rho$-corrected version for clarity (Fig.~\ref{fig:spacing_compare}b).

Interestingly, after correcting for the residual influence of $\rho$, the relationship between spacing and torus size $s_0$ follows a clear cubic law. Unlike the inverse dependence on $\rho$, which can be derived analytically, we do not yet have a theoretical explanation for this cubic dependence. Nevertheless, the fit is highly robust across simulations, suggesting that it reflects a genuine structural property of the model rather than a numerical artifact.

\section{Additional Firing Field Examples}\label{sec:FF}

\begin{figure}[h]
  \centering
  \includegraphics[width=\linewidth]{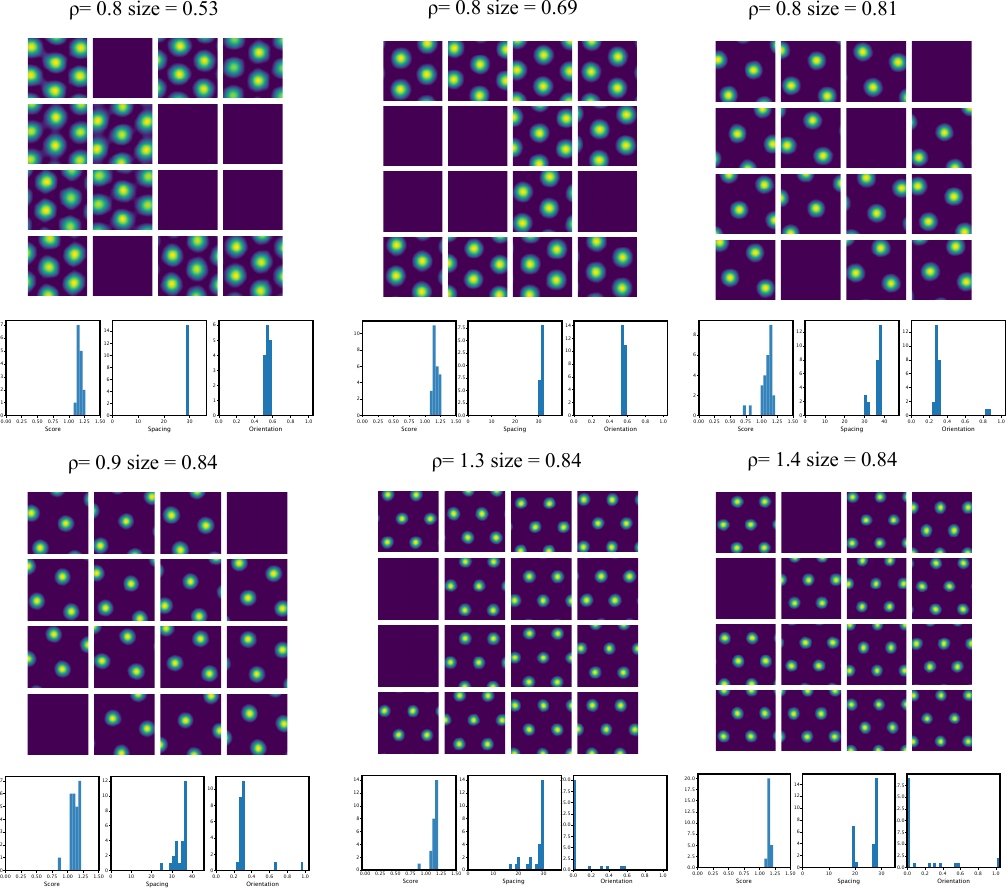}
    \caption{Additional firing map examples. }
  \label{fig:more_fields}
\end{figure}

To complement the main text, Fig.~\ref{fig:more_fields} presents additional examples of firing fields generated by TopoCN under different training regimes. These visualizations illustrate the variability across runs while consistently demonstrating the emergence of hexagonal structure under appropriate torus-size constraints.

\end{document}